\documentclass[tighten,apj]{emulateapj}
\usepackage{natbib,graphicx,amsmath,amsfonts,amssymb,subfigure,enumerate,appendix}
\usepackage[backref,breaklinks,colorlinks,allcolors=blue,backref=false]{hyperref}
\newcommand{\kms}{~km~s$^{-1}$}

\newcommand{\hi}{\ion{H}{1}}
\defcitealias{Bradford:2015km}{I}

\def\NISOLATEDBTF{930}
\def\NLMISOLATEDBTF{271}
\def\NHMISOLATEDBTF{659}
\def\HIGHMASSPERC{71}
\def\LOWMASSPERC{29}
\def\nliteraturemeasures{25}
\def\minliteratureslope{3.0}
\def\maxliteratureslope{4.3}
\def\flatmedslope{3.9}
\def\maxmedslope{3.4}
\def\wfiftymedslope{3.9}
\def\wtwentymedslope{3.3}

\def\ctfslope{-8.50}
\def\ctfslopeerr{0.13}
\def\ctfconst{-0.12}
\def\ctfconsterr{0.28}
\def\ctfscat{0.81}
\def\ctfscaterr{0.02}
\def\ctfpearsonr{-0.90}
\def\stfslope{4.16}
\def\stfslopeerr{0.06}
\def\stfconst{0.82}
\def\stfconsterr{0.13}
\def\stfscat{0.32}
\def\stfscaterr{0.01}
\def\stfpearsonr{0.90}
\def\gtfslope{2.57}
\def\gtfslopeerr{0.05}
\def\gtfconst{4.32}
\def\gtfconsterr{0.10}
\def\gtfscat{0.28}
\def\gtfscaterr{0.01}
\def\gtfpearsonr{0.87}
\def\btfslope{3.24}
\def\btfslopeerr{0.05}
\def\btfconst{3.21}
\def\btfconsterr{0.10}
\def\btfscat{0.25}
\def\btfscaterr{0.01}
\def\btfpearsonr{0.92}
\def\outstateB{For (B), the slope changes by $-0.6$\%, the scatter changes by $8.0$\%, and the zero-point changes by $-0.3$\%.~}
\def\outstateC{For (C), the slope changes by $0.3$\%, the scatter changes by $0.0$\%, and the zero-point changes by $-1.2$\%.~}
\def\outstateD{For (D), the slope changes by $-1.9$\%, the scatter changes by $0.0$\%, and the zero-point changes by $5.0$\%.~}
\def\outstateE{For (E), the slope changes by $-9.0$\%, the scatter changes by $4.0$\%, and the zero-point changes by $19.3$\%.~}
\def\outstateF{For (F), the slope changes by $9.0$\%, the scatter changes by $-44.0$\%, and the zero-point changes by $-21.5$\%.~}
\def\outstateG{For (G), the slope changes by $-4.6$\%, the scatter changes by $0.0$\%, and the zero-point changes by $13.7$\%.~}
\def\outstateH{For (H), the slope changes by $-13.9$\%, the scatter changes by $12.0$\%, and the zero-point changes by $34.6$\%.~}
\def\outstateI{For (I), the slope changes by $-18.5$\%, the scatter changes by $-8.0$\%, and the zero-point changes by $44.9$\%.~}
\def\outstateJ{For (J), the slope changes by $8.0$\%, the scatter changes by $-28.0$\%, and the zero-point changes by $-18.4$\%.~}
\def\outstateK{For (K), the slope changes by $0.3$\%, the scatter changes by $-8.0$\%, and the zero-point changes by $-0.9$\%.~}
\def\outstateL{For (L), the slope changes by $1.5$\%, the scatter changes by $52.0$\%, and the zero-point changes by $-5.0$\%.~}
\def\outstateM{For (M), the slope changes by $-3.7$\%, the scatter changes by $8.0$\%, and the zero-point changes by $9.3$\%.~}
\def\outstateN{For (N), the slope changes by $0.9$\%, the scatter changes by $-36.0$\%, and the zero-point changes by $-2.2$\%.~}
\def\outstateO{For (O), the slope changes by $-6.5$\%, the scatter changes by $64.0$\%, and the zero-point changes by $11.8$\%.~}
\def\minsystslope{2.64}
\def\maxsystslope{3.53}
\def\minsystscatter{0.14}
\def\maxsystscatter{0.41}

\def\minbtfbaryonicmass{7.4}
\def\maxbtfbaryonicmass{11.3}

\def\nsteepgalaxies{729}
\def\slfwdolsslope{3.0}
\def\slrevolsslope{3.7}
\def\slbisolsslope{3.3}
\def\slortholsslope{3.6}
\def\slredmajslope{3.3}
\def\slmeanolsslope{3.3}
\def\sldiffslope{0.37}
\def\steepcutoff{0.17}
\def\nnonisogal{1308}
\def\nonisogallow{316}
\def\nedgeon{76}

\begin{document}
\submitted{2016 Feb 8}
\revised{2016 Aug 3}
\accepted{2016 Aug 5}

\title{A Slippery Slope: Systematic Uncertainties in the Line Width Baryonic Tully--Fisher Relation}

\author{Jeremy D. Bradford \altaffilmark{1,2}, Marla C. Geha \altaffilmark{1}, Frank C. van den Bosch \altaffilmark{1}}

\affil{$^1$Astronomy Department, Yale University, New Haven, CT~06520, USA; \href{mailto:jeremy.bradford@yale.edu}{jeremy.bradford@yale.edu}}

\altaffiltext{2}{NSF Graduate Research Fellow}

\begin{abstract}
\renewcommand{\thefootnote}{\fnsymbol{footnote}}

The baryonic Tully--Fisher relation (BTFR) is both a valuable observational tool and a critical test of galaxy formation theory. We explore the systematic uncertainty in the slope and the scatter of the observed line width BTFR utilizing homogeneously measured, unresolved \hi~observations for \NISOLATEDBTF~isolated galaxies. We measure a fiducial relation of $\log_{10}{M_{\rm baryon}} = \btfslope\log_{10}{V_{\rm rot}}~+~\btfconst$ with observed scatter of \btfscat~dex over a baryonic mass range of $10^{\minbtfbaryonicmass}$ to $10^{\maxbtfbaryonicmass}$M$_{\odot}$ where $V_{\rm rot}$ is measured from 20\% \hi~line widths. We then conservatively vary the definitions of $M_{\rm baryon}$ and $V_{\rm rot}$, the sample definition and the linear fitting algorithm. We obtain slopes ranging from \minsystslope~to \maxsystslope~and scatter measurements ranging from \minsystscatter~to \maxsystscatter~dex, indicating a significant systematic uncertainty of 0.25 in the BTFR slope derived from unresolved \hi~line widths. We next compare our fiducial slope to literature measurements, where reported slopes range from \minliteratureslope~to \maxliteratureslope~and scatter is either unmeasured, immeasurable or as large as 0.4~dex. Measurements derived from unresolved \hi~line widths tend to produce slopes of \wtwentymedslope, while measurements derived strictly from resolved asymptotic rotation velocities tend to produce slopes of \flatmedslope. The single largest factor affecting the BTFR slope is the definition of rotation velocity. The sample definition, the mass range and the linear fitting algorithm also significantly affect the measured BTFR. We find that galaxies in our sample with $V_{\rm rot} < 100$\kms~are consistent with the line width BTFR of more massive galaxies, but these galaxies drive most of the observed scatter. It is critical when comparing predictions to an observed BTFR that the rotation velocity definition, the sample selection and the fitting algorithm are similarly defined. We recommend direct statistical comparisons between data sets with commensurable properties as opposed to simply comparing BTFR power-law fits.

\end{abstract}


\maketitle

\section{Introduction}
\label{sec_intro}

The Tully--Fisher relation (TFR) is an observed relation between the rotation velocity of disk galaxies and the total amount of observed luminous material \citep{Tully:1977wu}. The TFR is the disk-galaxy scaling relationship with the least scatter and therefore may be the most fundamental \citep[e.g.,][]{Courteau:2007ek, AvilaReese:2008gz}. The TFR is classically used as a distance measurement for spiral galaxies \citep[e.g.,][etc.]{Opik:1922kk, Roberts:1969bx, Tully:1977wu, Tully:1985kb, Courteau:1993bs, Courtois:2009ix, Tully:2012eg, Sorce:2013hn, Sorce:2014by,Neill:2014fr} and to provide a measure of the fraction of luminous baryons that have condensed out of and formed into a galactic disk \citep[e.g.,][]{Cole:1994vb, Dalcanton:1997uk, Mo:2000bj, vandenBosch:2000iw}.

In low-mass galaxies with rotational velocities below $\sim100$\kms, \textit{observable} baryons are often dominated by atomic gas \citep[e.g.,][]{Huang:2012gk, Bradford:2015km}. Thus many, but not all, low-mass galaxies lie at lower luminosities (larger velocities) relative to the TFR of high-mass galaxies \citep[]{Freeman:1999un, Walker:1999jb, McGaugh:2000hx}. If atomic gas masses are included in the accounting of baryonic matter, the ``kink" in the local TFR is mostly corrected and the scatter in the relation is minimized \citep{Stark:2009ks, McGaugh:2012ev}. In terms of the low-mass end of spiral galaxy populations and galaxy formation theory, the classic TFR has been largely supplanted by the baryonic Tully--Fisher relation (BTFR). Therefore, for the purpose of our study, we focus almost entirely on the BTFR and the relevant literature. Any galactic baryonic mass ($M_{\rm baryon}$) measurement that scales as a function of a rotation velocity measurement ($V_{\rm rot}$) is generally referred to as the BTFR, regardless of the sample, the rotation velocity definition, the mass range, or the fitting algorithm used to measure the relation. 

As an empirical relationship, the astrophysical applications of the BTFR are wide ranging. It is often used to estimate any baryonic component of a galaxy that is unknown \citep[e.g.,][]{McGaugh:2015eu, Sorce:2015kq, Wang:2015gh}. It has also been used to examine the deviations of tidal dwarfs from the BTFR \citep{Lelli:2015wc}, to compare the line-of-sight velocities of \ion{Mg}{2} absorbers to the rotation velocities of stellar disks \citep{DiamondStanic:2015th}, to compare giant disk galaxies to ``regular" galaxies on the BTFR \citep{Courtois:2015iqa}, and even to detect the effect of civilizations on their host galaxy light emission \citep{Zackrisson:2015fg}. 

Above all, the BTFR is a vital metric used to quantify the success of a particular galaxy formation model or simulation. Within the $\Lambda$CDM framework, the BTFR has its origin in a combination of the virial relation for dark matter halos, $M_{\rm vir} \propto V_{\rm vir}^3$, the baryonic mass fraction of dark matter halos, $M_{\rm baryon}/M_{\rm vir}$, and the relation between the observed rotation velocity of a disk-galaxy and the halo virial velocity, $V_{\rm obs}/V_{\rm vir}$ \citep[e.g.,][]{Mo:1998hg, vandenBosch:2000iw, Dutton:2007he}. The latter relates to the self-gravity of the disk and the amount of contraction or expansion a halo undergoes during the disk formation process. 

Consequently, the BTFR is an important testbed for the growth and evolution of baryonic disks within dark matter halos. Simultaneously matching the slope, zero-point, and scatter of the observed BTFR (or the classic TFR) is a standard test of both semi-analytic models of galaxy formation \citep[e.g.,][]{Cole:2002fl, Dutton:2007he, Dutton:2009dn, Dutton:2012jc, Desmond:2015gr} and hydro-dynamic simulations \citep[e.g.,][]{Navarro:2000bt, Governato:2007dq, Agertz:2010dz, Brook:2012ht, Vogelsberger:2014tw, Chan:2015kl}. 

$\Lambda$CDM predictions result in a BTFR slope between 3 and 4 with varying but significant amounts of scatter. The specific prediction is dependent on the model, feedback implementation, rotation velocity definition, baryonic tracer used to measure $V_{\rm rot}$, sample size, and underlying galaxy distribution \citep{Dutton:2012jc,Brook:2015fz, DiCintio:2016tj, SantosSantos:2015em, Brook:2016gh, Sorce:2016jj}. Observational results vary, but several independent BTFR studies are often cited as being consistent with the $\Lambda$CDM paradigm while others are not (see \S\,\ref{subsec_lit}). It is unclear from the present literature whether this tension is due to the predictions, the methods used to calibrate the BTFR, or the physical measurements that go into the BTFR. Indeed, the BTFR becomes something of a fine-tuning problem for $\Lambda$CDM, where details of simulated galaxies' rotation curves are often subsumed into the uncertainty of sub-grid physics or modeled feedback processes. In the end, predicted rotation velocities are often measured at some fixed characteristic radius or \hi~surface density and then compared to observed BTFRs calibrated with a variety of rotation velocities and galaxy sample definitions.

In contrast to $\Lambda$CDM, modified Newtonian dynamics (MOND) strictly predicts a BTFR slope of 4 with no scatter \citep{McGaugh:2011fq}. Since the first BTFR fit published in \citet{McGaugh:2000hx}, many subsequent observational fits have been published that are consistent with $M_{\rm baryon} \propto V_{\rm rot}^4$ with little to no scatter  over a large range of baryonic masses \citep{Verheijen:2001hp, McGaugh:2005bc, Stark:2009ks, Trachternach:2009ff, McGaugh:2012ev, Bottema:2015ji, McGaugh:2015eu, Lelli:2016ed}. However, the observed asymptotic velocity of galaxies becomes more difficult and often impossible to measure at the lowest and highest baryonic masses \citep{Verheijen:2001hp}. Because resolved \hi~data is difficult to obtain and many rotation curves do not probe the flattening velocity, there are roughly 200 galaxies that have been observed to have \textit{strictly} asymptotic rotation curves that pass the quality cuts of the resolved studies mentioned here (see \S\,\ref{subsec_lit}). 

Because the resolved asymptotic velocity is either difficult or impossible to measure in most galaxies, we hope to take advantage of the rotation velocities and baryonic masses that are derived from large, single-dish, \hi~surveys. These large data sets are dominated by gas-rich galaxies, which mitigate the uncertainty and effect of systematics in stellar mass calculations when calibrating the BTFR. These large data sets are also helpful when building up a statistical consensus for comparisons to predictions of galaxy formation. Unresolved line widths, however, are systematically affected by the fact that we do not know precisely what rotation velocity has been measured. If we are to take advantage of the wealth of unresolved data, a careful examination of these systematics is needed.

We revisit the BTFR derived from \hi~line widths in the context of a homogeneously measured and publicly available data set that includes a significantly larger number of isolated low-mass galaxies than previously available. We are motivated by the pervasive use of the BTFR in the recent literature and the assumptions that go into measuring rotation velocities and baryonic masses, specifically at low masses. These assumptions may drive significant systematic uncertainty in the BTFR. Indeed, it is well-known that the slope and scatter of the observed TFR are strongly dependent on color, distance measurement, morphology, sample selection, photometric band, rotational velocity probe, and environment \citep[e.g.,][]{Courteau:1997ic, Giovanelli:1997fc, Matthews:1998gq, Tully:2000be, Bell:2001hv, Verheijen:2001hp, Kannappan:2002ic, Bedregal:2006jp, Saintonge:2010gg}. If we are to correctly interpret large, unresolved statistical samples - it is critical that we understand exactly what we are measuring with the BTFR and subsequent observations that rely on line widths derived from unresolved HI. Our goal is to determine how much uncertainty exists in the line width derived BTFR, what drives this uncertainty, and what the implications are when applying unresolved \hi~observations of the BTFR as a litmus test for galaxy formation.

This paper is outlined as follows: In \S\,\ref{sec_data}, we describe a homogeneous data set of isolated galaxies and derive a fiducial BTFR based on these data in \S\,\ref{subsec_btf}. In \S\,\ref{subsubsec_btf_scatslope}, we use this data set to explore systematic uncertainties in the slope and the scatter of the BTFR. In \S\,\ref{subsec_lit}, we compare our fiducial BTFR to similar measurements in the literature. In \S\,\ref{sec_discussion}, we discuss our results. In comparing between the observed BTFR and the predicted BTFR, we conclude that the measurement methods, sample selection and fitting algorithms must be similar in order for such comparison to be meaningful. Throughout this paper, we assume a $\Lambda$CDM cosmology with $\Omega_{\rm m} = 0.3$, $\Omega_{\rm \Lambda} = 0.7$ and $H_0 = 70 \rm km~s^{-1}~Mpc^{-1}$ (i.e., $h = 0.7$). All data used to produce the results and figures below are available at the lead author's personal website\footnote{\url{http://www.astro.yale.edu/jdbradford/}}.

\section{Data and Methods} 
\label{sec_data}

In this section, we provide a brief description of the data set, including the isolation criteria of our galaxies and the observed and derived parameters.

\subsection{Galaxy Catalog}
\label{subsec_nsa}

Our BTFR data set is based on the NASA-Sloan Atlas\footnote{\url{http://www.nsatlas.org}} (NSA), which is a re-reduction of the Sloan Digital Sky Survey (SDSS) DR-8 \citep{Aihara:2011kj} and is optimized for low-redshift galaxies with an improved background subtraction technique \citep{Blanton:2011dv}. We select galaxies with single-dish \hi~observations. All analysis is performed on isolated galaxies to minimize effects of environment \citep{Haynes:1984fs}. The isolation criteria for our sample is defined below. This catalog has been described in \citet{Bradford:2015km}, hereafter referred to as Paper~\citetalias{Bradford:2015km}. 

Stellar masses ($M_*$) are calculated using the \texttt{kcorrect} software of \citet{Blanton:2007kl} with a \citet{Chabrier:2003ki} IMF using GALEX and SDSS photometric bands. We assume a systematic 0.2 dex uncertainty in $M_*$ (see Paper~\citetalias{Bradford:2015km}). Luminosity distances ($D$) are computed with peculiar velocity corrected recession velocities using the model of \citet{Willick:1997gg}. All galaxies must have $g$, $r$ and $i$ photometry to ensure accurate stellar masses; peculiar velocity corrected redshifts greater than 0.002 to ensure accurate distances; and S\'ersic minor-to-major axis ratios less than 0.65 to ensure the most accurate inclination-corrected \hi~line width velocities. In Paper~\citetalias{Bradford:2015km}, we found that the NSA catalog surface brightness measuring algorithm occasionally interprets face-on barred galaxies as edge-on galaxies. These barred galaxies artificially inflate the scatter in the BTFR due to incorrect inclination measurements, therefore we remove any barred galaxies that have been identified as such in the Galaxy Zoo project \citep{Hoyle:2011bk}. 

Single-dish \hi~observations were obtained using the 305\,m Arecibo Telescope (AO) and the 100\,m Green Bank Telescope (GBT) between Spring 2005 and Fall 2014. All spectra are smoothed to a 5\,\kms~resolution. Our \hi~data reduction technique is fully automated and is described in detail in Paper~\citetalias{Bradford:2015km}. We measure velocity widths at both 20\% of the peak flux ($W20$) and at 50\% the peak flux ($W50$). We compute inclination-corrected rotation velocities and gas masses as in \S\ \ref{subsec_est}. 

We combine these \hi~observations with the code-1 sources from the ALFALFA survey 40\% data release that overlap with our isolated galaxy sample. The ALFALFA survey is a blind, flux-limited survey that overlaps much of the SDSS footprint \citep{Haynes:2011en}. In order to ensure homogeneity of \hi~parameters and to obtain 20\% line widths for all galaxies, we re-analyze the ALFALFA spectra by running the data through our pipeline. See Paper~\citetalias{Bradford:2015km} for a comparison between our and ALFALFA's derived parameters. We note that the ALFALFA survey and our \hi~data set are significantly biased toward blue, star-forming, gas-rich galaxies.

Environmental processes can affect a galaxy's position on the BTFR. To minimize these effects, we select a sample of isolated galaxies to fit the BTFR as in Paper~\citetalias{Bradford:2015km}. For galaxies with $M_{*} < 10^{9.5} M_{\odot}$, we calculate the projected distance ($d_{\rm host}$) to the nearest galaxy with $M_* > 10^{10} M_{\odot}$ and $c\Delta z < 1000$\kms. Following from \citet{Geha:2012eu}, we select isolated low-mass galaxies with $d_{\rm host} > 1.5$~Mpc. For galaxies with $M_{*} > 10^{9.5} M_{\odot}$, we calculate the projected distance to the nearest galaxy that is more massive by 0.5 dex ($d_{\rm host, 0.5}$) and with $c\Delta z < 1000$\kms. Because few galaxies exist in our sample with $M_{*} > 10^{10.5} M_{\odot}$, we also compute the fifth nearest neighbor surface density ($\Sigma_5$) for these high-mass galaxies. Following from Paper~\citetalias{Bradford:2015km}, we select isolated high-mass galaxies with $d_{\rm host, 0.5} > 1.5$~Mpc and $\Sigma_5 < 1~\rm Mpc^{-2}$.

Our final BTFR sample consists of \NISOLATEDBTF~galaxies that are isolated and have well-measured stellar masses, atomic gas masses, and \hi~line width rotation velocities. This sample is by no means complete. Our data set is heavily biased toward gas-rich galaxies at all masses and is better sampled at the more massive end of the BTFR (Paper I). We examine this bias in our analysis below.

\subsection{Mass and Velocity Estimates}
\label{subsec_est}

We calculate \hi~masses using the standard formula,

\begin{equation}
M_{\rm HI} = 2.356\times10^5 [M_{\odot}] \left(\frac{D}{\rm Mpc}\right)^2 \frac{S_{21}}{\rm Jy~km~s^{-1}},
\end{equation}

\noindent where D is defined as the distance to the galaxy (see \S\ \ref{subsec_nsa}) and $S_{21}$ is the integrated \hi~flux density \citep[][p. 309]{Roberts:1975uj}. Unless explicitly noted, we calculate the total atomic gas mass as,

\begin{equation}
M_{\rm gas} = 1.4M_{\rm HI},
\label{eq_M_gas}
\end{equation}

\noindent to correct for a solar helium (He) abundance \citep{Arnett:1996vt}. Therefore, we calculate total baryonic mass as,

\begin{equation}
M_{\rm baryon} = M_* + M_{\rm gas}.
\label{eq_M_baryon}
\end{equation}
 
We calculate rotation velocities by correcting for inclination and redshift broadening. We compute inclinations as,

\begin{equation}
\sin{i} = \sqrt{\frac{1 - (b/a)^2}{1 - q_0^2}},
\end{equation}

\noindent where $b/a$ is the observed minor-to-major axis ratio derived from the NSA catalog and $q_0$ is the intrinsic minor-to-major axis ratio. We assume that $q_0 = 0.2$ in our fiducial analysis \citep{Hubble:1926be, Haynes:1984fs}. Many studies set $q_0 \sim 0.2$ (see \citeauthor{Yuan:2004kr} for an excellent discussion of the history of $q_0$).  

The value of $q_0$ has been shown to be larger for many low-mass galaxies with irregular morphologies \citep{Verheijen:2001gr, Roychowdhury:2010kt, SanchezJanssen:2010fw, Leaman:2012di, Roychowdhury:2013gz}. For a brief examination of what such a systematic change in disk thickness would do to the resulting BTFR, we calculate $q_0$ as a simple linear function that varies from 0.15 for the most massive galaxies to 0.5 for the least massive galaxies. We apply this linear function to Equation \ref{eq_v_WX} and find that this has almost no effect on the BTFR so we forgo further analysis and assume $q_0 = 0.2$ for the rest of this work.

We compute inclination- and redshift-broadening- corrected rotation velocities as,

\begin{equation}
V_{X,i} = \frac{W_{X}}{2} \frac{1}{\sin{i}} \frac{1}{(1 + z)},
\label{eq_v_WX}
\end{equation}

\noindent with $X = W20$ or $X = W50$ corresponding to either the 20\% or 50\% line width percentage, respectively, and $z$ is the observed redshift \citep{Haynes:1984el, Giovanelli:1997jz, Springob:2005db}. We compute errors on \hi~line widths using the Monte Carlo error estimation procedure described in Paper~\citetalias{Bradford:2015km}. We remove galaxies from our sample when $V_{W50,i}$ velocity errors are greater than 50\% of the measured value. This cut on $V_{W50,i}$ is intended to remove galaxies where the 50\% line width measurement has failed. A failed measurement is not always due to the noise figure but also due to the observed shape of the \hi~emission line.

\begin{figure*}[t!]
\epsscale{1.22}
\plotone{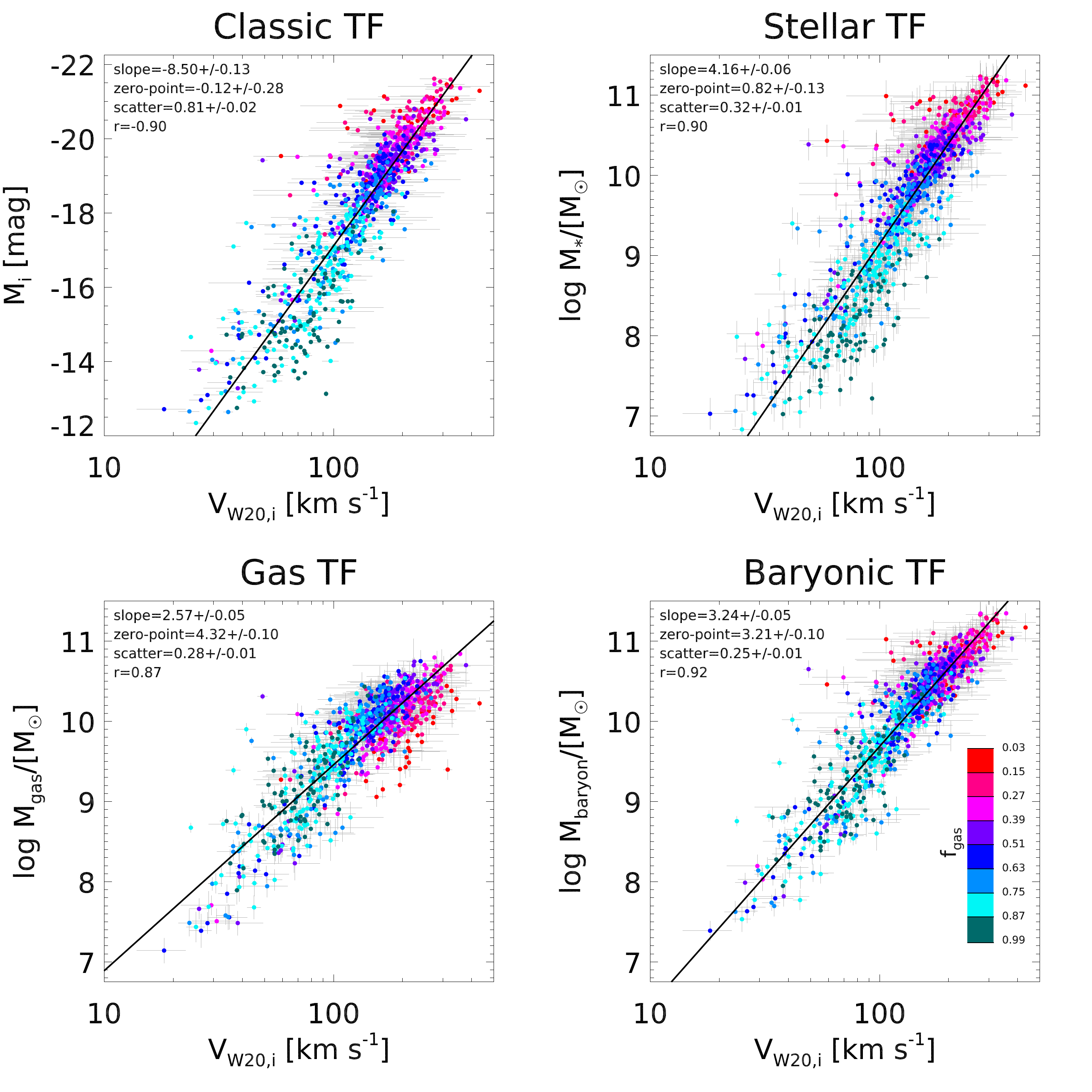}
\caption{The fiducial classic (top left), stellar (top right), atomic gas-only (bottom left), and baryonic (bottom right) Tully--Fisher relations. The slope, zero-point, scatter, and Pearson r measurements are listed in the top left of each panel. Data points are color-coded by gas fraction according to the legend in the bottom right panel. As expected, the slight downturn for some low-mass galaxies in the stellar relation is corrected when atomic gas masses are included in the y axis. The scatter is minimized in the baryonic relation. The range of each mass and rotation velocity axis are the same for ease of comparison between relations. The data used to create this figure are available.}
\label{fig_btf_primary}
\end{figure*}

\section{Results}
\label{sec_results}

We perform a fiducial calibration of the BTFR to serve as an anchor for the rest of our analysis and to compare against the classic, stellar, and gas-only TFRs. This fiducial calibration is not necessarily the ``correct" one. These fiducial models simply act as a point of reference. We then analyze the impact of systematic uncertainties, sample selections and various assumptions on the slope and scatter of the BTFR. Finally, we examine many BTFR measurements in the literature with a focus on how rotational velocities are defined for each study.

\subsection{Fitting the Linear Relation}
\label{subsec_fit_lin}
Throughout this work, we assume the BTFR is a linear relation in log space between $V_{\rm rot}$ (derived here from \hi~line widths) and $M_{\rm baryon}$ across all decades of mass with an intrinsic scatter in $M_{\rm baryon}$ as a function of rotation velocity:

\begin{equation}
\log_{10}{M_{\rm baryon}} = \alpha + \beta\log_{10}V_{\rm rot} \pm \epsilon,
\label{eq_btfr}
\end{equation}

\noindent where $\alpha$ is the zero-point, $\beta$ is the slope, and $\epsilon$ is the observed scatter about the relation in $\log_{10}{M_{\rm baryon}}$. 

If the scatter is not modeled in the linear fit, then the galaxies with the best precision will dominate the model \citep{Kelly:2007bv, Sereno:2016he}. This is because most fitting algorithms without scatter and errors in both dimensions give more weight to data points with the smallest uncertainties in the dependent variable \citep{Weiner:2006bv}. Since our data are more heavily sampled at the high-mass end and high-mass galaxies tend to have higher signal-to-noise ratios (S/N) than low-mass galaxies, performing linear regression without considering scatter and errors in both dimensions can severely impact the resulting slope and zero-point of the BTFR \citep[e.g.,][]{Willick:1997gg, Courteau:1997ic, Saintonge:2010gg, McGaugh:2012ev}. 

Unless specifically mentioned, we fit all linear relations using the \citet{Kelly:2007bv} Bayesian linear regression fitting algorithm with a minimum of 5000 iterations. This algorithm provides a generative model of the BTFR as it accounts for uncertainty in both $M_{\rm baryon}$ and $V_{\rm rot}$, correlated uncertainty in both $M_{\rm baryon}$ and $V_{\rm rot}$, uncertainties that may vary in either $V_{\rm rot}$ or $M_{\rm baryon}$ and intrinsic scatter in the BTFR. The \citet{Kelly:2007bv} algorithm overcomes heteroscedasticity and sampling (Malmquist) bias by generating a model of the ``true" independent variables that consists of weighted sums of Gaussian functions. This algorithm follows structural equation modeling, meaning the algorithm generates models that are valid given the observed data using a maximum likelihood approach. Modeling the intrinsic scatter in the BTFR is critical because the observed scatter may be partly due to parameters that are not included in the data we are fitting \citep{Hogg:2010wk}. Scatter in galaxy scaling relations is driven by the details of galaxy formation \citep{Dutton:2007he} (e.g., velocity dispersion, feedback efficiency, baryonic surface density, adiabatic contraction, formation history, etc.). We also explore several alternative algorithms that are commonly used to fit the BTFR in \S\,\ref{subsubsec_btf_scatslope}.

\subsection{The Fiducial Tully--Fisher Relations}
\label{subsec_btf}

\begin{deluxetable}{lrrrr}
\tabletypesize{\tiny}
\tablecaption{Fiducial Tully--Fisher Relations}
\tablewidth{0pt}
\tablehead{
\colhead{TFR} &
\colhead{Slope} &
\colhead{Zero-point} &
\colhead{Scatter} &
\colhead{Pearson r}}
\startdata
Classic & $\ctfslope \pm \ctfslopeerr$ & $\ctfconst \pm \ctfconsterr$ & $\ctfscat \pm \ctfscaterr$ & \ctfpearsonr \\
Stellar & $\stfslope \pm \stfslopeerr$ & $\stfconst \pm \stfconsterr$ & $\stfscat \pm \stfscaterr$ & \stfpearsonr \\
Gas & $\gtfslope \pm \gtfslopeerr$ & $\gtfconst \pm \gtfconsterr$ & $\gtfscat \pm \gtfscaterr$ & \gtfpearsonr \\
Baryonic & $\btfslope \pm \btfslopeerr$ & $\btfconst \pm \btfconsterr$ & $\btfscat \pm \btfscaterr$ & \btfpearsonr
\enddata
\tablecomments{Fiducial Tully--Fisher relations (e.g., equation \ref{eq_btfr}) presented in Figure \ref{fig_btf_primary}, as described in \S\ \ref{subsec_btf}.}
\label{tbl_fid_tfr_fits}
\end{deluxetable}

We fit the classic, stellar, atomic gas-only, and baryonic TFRs based on the galaxy sample and calculations described in \S\ \ref{sec_data} and the fitting method described in \S\ \ref{subsec_fit_lin}. Atomic gas mass is calculated with solar He abundance as in Equation \ref{eq_M_gas}. The $M_{\rm baryon}$ of our fiducial BTFR calibration is calculated as in Equation \ref{eq_M_baryon}. The rest of the relations follow the same linear form as Equation \ref{eq_btfr} but with $i$-band absolute magnitudes, $M_*$, and $M_{\rm gas}$ as the dependent variables. The $V_{\rm rot}$ of our fiducial calibration is calculated as in Equation \ref{eq_v_WX}, where $X = 20$ ($V_{W20,i}$). 

The exploration of various line width definitions and measurement algorithms with relation to the TFR has been studied extensively over the past $\sim 40$ years \citep[e.g.,][]{Roberts:1978fl, Mould:1980hg, Aaronson:1982cx, Bottinelli:1982tt, Bottinelli:1983ve, Haynes:1984el, Bicay:1986br, Tully:1985kb, Giovanelli:1997fc, Haynes:1999gl}. A thorough review of the topic is beyond the scope of this paper, but we find varying motivations for the use of 20\% or 50\% line widths throughout the BTFR and TFR literature. We use the inclination-corrected 20\% line widths here, rather than the 50\% line widths, for several reasons. We find that at low masses, our measurement algorithm is more sensitive to noise features at the 50\% line width than the 20\% line width. This is because low-mass galaxies' \hi~profiles can be asymmetric and these \hi~emission lines tend to include noise features that might be misinterpreted as single- or double-peaked profiles. Noise features near the center of \hi~emission lines can dramatically affect the value of the measured peak flux. While methods exist to mitigate this effect, such as taking the median peak flux for multi-peaked emission lines \citep[e.g.,][]{Springob:2005db}, we find that these methods increase both the number of assumptions made about the resulting \hi~parameters and the amount of human intervention required when measuring these parameters. The TFRs measured here with $V_{W20,i}$ are most consistent with similar measurements in the literature. $V_{W20,i}$ also minimizes residuals at the low-mass end of the BTFR. We assume that the best line width for any BTFR study is strongly dependent on the \hi~measurement algorithm employed, the bias of the galaxy sample, and the mass-regime of the galaxy sample studied. We also examine the use of 50\% line widths in \S\ \ref{subsubsec_btf_scatslope}.

\begin{figure*}[t!]
\epsscale{1.2}
\plotone{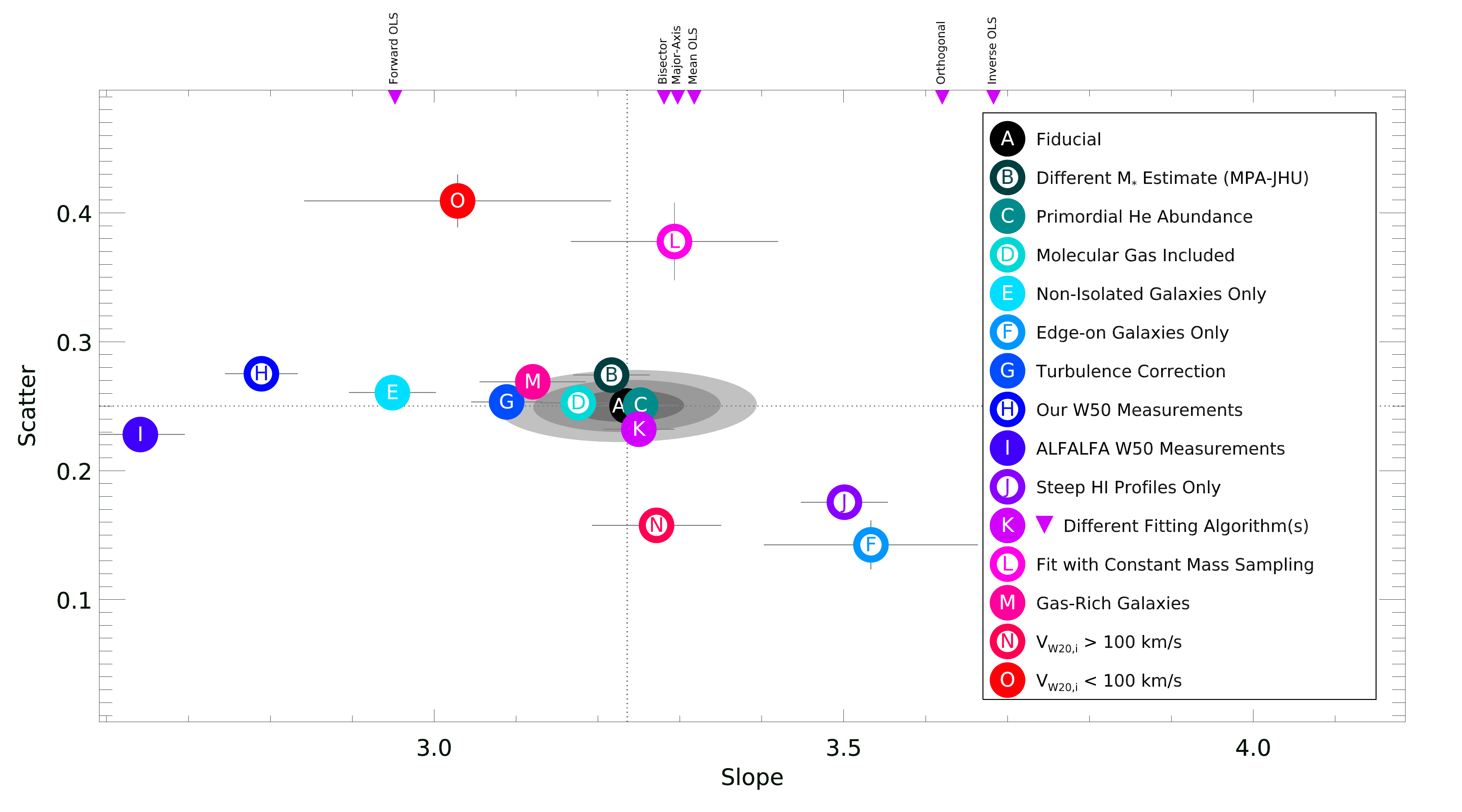}
\caption{The slope and scatter for each systematic uncertainty, assumption, and selection effect (i.e., ``systematic") typically made when measuring the BTFR. Each data point represents the systematic listed in the legend and in \S\ \ref{subsubsec_btf_scatslope}. The figure is centered on our fiducial measurement (A). For comparison with the other BTFR measurements, we plot $1\sigma$, $2\sigma$ and $3\sigma$ uncertainty ellipses for the fiducial measurement (gray shaded regions). Error bars that are not visible are smaller than the plotted data point. We plot the fitting results from \texttt{SIXLIN} as six downward-pointing triangles on the top axis of the figure since these measurements do not include an estimate for scatter. As discussed in item (K) in \S\ \ref{subsubsec_btf_scatslope}, we do not include these six values in our reported slope measurements because these measurements do not account for uncertainty in both x and y and they are not fit with intrinsic scatter. The rotation velocity measurement method has the largest effect on the BTFR slope; the mass range of the selected sample has the largest effect on the scatter. For a brief description of each data point's sample size and underlying sample distribution, see Table \ref{tbl_btf_syst}.}
\label{fig_btf_syst}
\end{figure*}

We elect not to correct $V_{W20,i}$ for turbulence in our fiducial calibration of the BTFR. If we assume the BTFR is simply an observed relation, then subtracting a flat turbulent velocity correction from all \hi~line widths may create more problems than it solves by systematically shifting galaxies to smaller rotational velocities. This correction can also be dependent on the line width measurement algorithm. Non-rotational motion can be significant relative to rotational motion in low-mass galaxies. This non-rotational motion may probe the gravitational potential of galaxies that are partly dispersion supported \citep[e.g.,][]{Kassin:2012kz}. Indeed, including dispersion in the kinematic axis of the TFR has been shown to fix departures of low-mass galaxies from the high-mass TFR at higher redshifts \citep[e.g.,][]{Simons:2015ux}. We discuss the effect of turbulence and many other assumptions and systematics in \S\ \ref{subsubsec_btf_scatslope}. For now, we proceed with this as our fiducial choice of galaxy parameters. 

We present four TFRs in Figure \ref{fig_btf_primary}. Each fit is listed in Table \ref{tbl_fid_tfr_fits}. The data points in each panel of Figure \ref{fig_btf_primary} are color-coded by atomic gas fraction ($f_{\rm gas} = M_{\rm gas}/M_{\rm baryon}$) with the gas fractions increasing from red (mostly high-mass galaxies) to green (mostly low-mass galaxies; also see Figure 3 in Paper~\citetalias{Bradford:2015km}). The relation between $i$-band absolute magnitude and the log of the stellar mass is roughly linear, so the classic and stellar TFRs are similar in steepness and relative scatter. The slope of the stellar mass TFR is close to the maximum BTFR slope that has been measured in the literature, while the atomic gas TF relation is close to the minimum BTFR slope measured in the literature (see \S\ \ref{subsec_lit}). We note that one motivating factor for the calibration of our fiducial TFRs is to be consistent with the various TFRs in the literature, so it is not surprising that our measurements are consistent.

\subsection{Systematic Uncertainties in the BTFR Scatter and Slope}
\label{subsubsec_btf_scatslope}

We next explore how various assumptions, systematic uncertainties, and selection effects affect the fiducial BTFR fit (hereafter referred to only as systematics). Our goal is not necessarily to obtain the most correct answer for the fit of the BTFR but to examine how the observed BTFR fit changes when varying the assumptions that the typical observer will make. A complete analysis of our sample completeness and bias as well as an exploration of the effect of underestimated uncertainties on the observed scatter in the BTFR are beyond the scope of this work. 

Motivation for an analysis of systematics comes from the fact that our data set offers an opportunity to perform a new, relatively independent calibration of the BTFR. We have assembled a large sample of isolated galaxies that probe down to some of the lowest baryonic masses typically measured in an unresolved BTFR study (see Figure \ref{fig_lit_data}). Isolated galaxies are an excellent probe of the BTFR and mitigate some effects of bias and mass uncertainties in the BTFR, as low-mass isolated galaxies tend to be dominated by neutral gas and are therefore more easily observed and less effected by uncertainties in $M_*$ \citep{Stark:2009ks}. We also calibrate the BTFR with homogeneous re-measurements of both the SDSS and the ALFALFA survey (Paper I).

While we have considered other systematics in the BTFR not listed here, the systematics below have either the largest effect on the BTFR or are the most relevant to assumptions typically made in the literature. For each systematic, we plot the resulting slope and observed scatter of the BTFR in Figure \ref{fig_btf_syst}. We also plot the fiducial calibration against each subsequent fit in Figure \ref{fig_btf_syst_all}. While we focus our analysis on the slope and the scatter in the BTFR, we also list each relation's sample size and distribution, fit and a comment on the underlying sample in Table \ref{tbl_btf_syst} for reference and we plot each relation in Figure \ref{fig_btf_syst_all}. See also the discussion of \citet{Hall:2012hh} and the appendix of \citet{Gurovich:2010ha} where the authors discuss several systematics that contribute to uncertainties in the BTFR slope.

\begin{enumerate}[(A)] 

\item {{\em{Fiducial:}}} we begin with the fiducial calibration of the BTFR (bottom right panel, Figure~\ref{fig_btf_primary}), which is described in \S\,\ref{subsec_btf} and Table~\ref{tbl_fid_tfr_fits}. This BTFR slope is consistent with many literature measurements calibrated with \hi~line widths (see Figure \ref{fig_lit_data}). We discuss each systematic below relative to this fiducial measurement. We consider the BTFRs below to be consistent with the fiducial measurement if the slope or scatter are within the fiducial $3\sigma$ error ellipses plotted in Figure \ref{fig_btf_syst}.

\item {{\em{Different Stellar Mass Estimates:}}} varying the methods of estimating $M_*$ and its related uncertainties may affect the BTFR \citep{McGaugh:2005bc}. Gas dominated galaxies offer leverage over this problem as the stellar mass contribution to $M_{\rm baryon}$ is minimized \citep{Stark:2009ks}. Indeed, a BTFR that has been calibrated using gas dominated galaxies can recover the same stellar masses as stellar population synthesis models \citep{McGaugh:2015eu}. 

We are limited to the photometry available for this SDSS derived sample, so we compare our fiducial \texttt{kcorrect} stellar masses to those based on the MPA-JHU SDSS DR-8 stellar mass \citep{Kauffmann:2003iz, Brinchmann:2004hy, Tremonti:2004ed}. The MPA-JHU stellar mass estimates are also derived from the SDSS $ugriz$ photometry but assume a \citet{Kroupa:2001ki} IMF as opposed to the Chabrier IMF of the NSA catalog. We note that although several studies have suggested the IMF varies as a function of stellar mass \citep{Geha:2013vf}, recent work suggests that the IMF deviates from either a Chabrier or Kroupa IMF only at the extremes of the galaxy mass function, which are not covered in our data set \citep{Offner:2015uf}.

\outstateB we find that the MPA-JHU stellar mass estimates, which are consistently smaller at all masses than those of our fiducial model, slightly decrease the slope and increase the scatter of the BTFR. The most significant change in $M_{\rm baryon}$ is at the high-mass end, where galaxies' baryonic masses are no longer dominated by gas. This slope and scatter are consistent with our fiducial measurement.

We have also compared our fiducial model to stellar mass estimates based on 2MASS K-band photometry using a mass-to-light ratio of 0.6 \citep{McGaugh:2014kx, Papastergis:2016tl}. The average offset between our fiducial stellar masses and the K-band masses is $0.06$~dex. 520 of our fiducial sample have K-band photometry in the 2MASS XSC catalog. The BTFR measured with these stellar masses are consistent with our fiducial model. We attribute this to the fact that our galaxy sample is so gas-rich and therefore only the most high-mass galaxies are sensitive to stellar mass changes. Given our fitting algorithm, systematic variation in stellar mass over a small mass range should leave the resulting fit unaffected as observed.

\item {{\em{Primordial He Abundance:}}} typically, a solar abundance of helium is assumed when converting \hi~mass to total atomic gas mass ($1.4M_{\rm HI}$). However, best-fitting multipliers of 3 can be found in the literature when the \hi~mass multiplier is allowed to be free in the BTFR fit \citep{Pfenniger:2005jd}. When working with metal-poor, low-mass spiral galaxies, a primordial hydrogen abundance of $M_{\rm gas} = 1.33 M_{\rm HI}$ might be more appropriate \citep{McGaugh:2012ev}. While not all of our galaxies are metal-poor, this allows us to examine the effect of assuming a primordial He abundance. \outstateC We find a primordial He abundance has little effect on the BTFR; the slope and scatter are consistent with our fiducial measurement. Although we have no evidence to support increasing the \hi~multiplier, increasing the multiplier to 3 decreases the slope to 3.0 and does not change the scatter in the relation.

\begin{figure*}[t!]
\epsscale{1.2}
\plotone{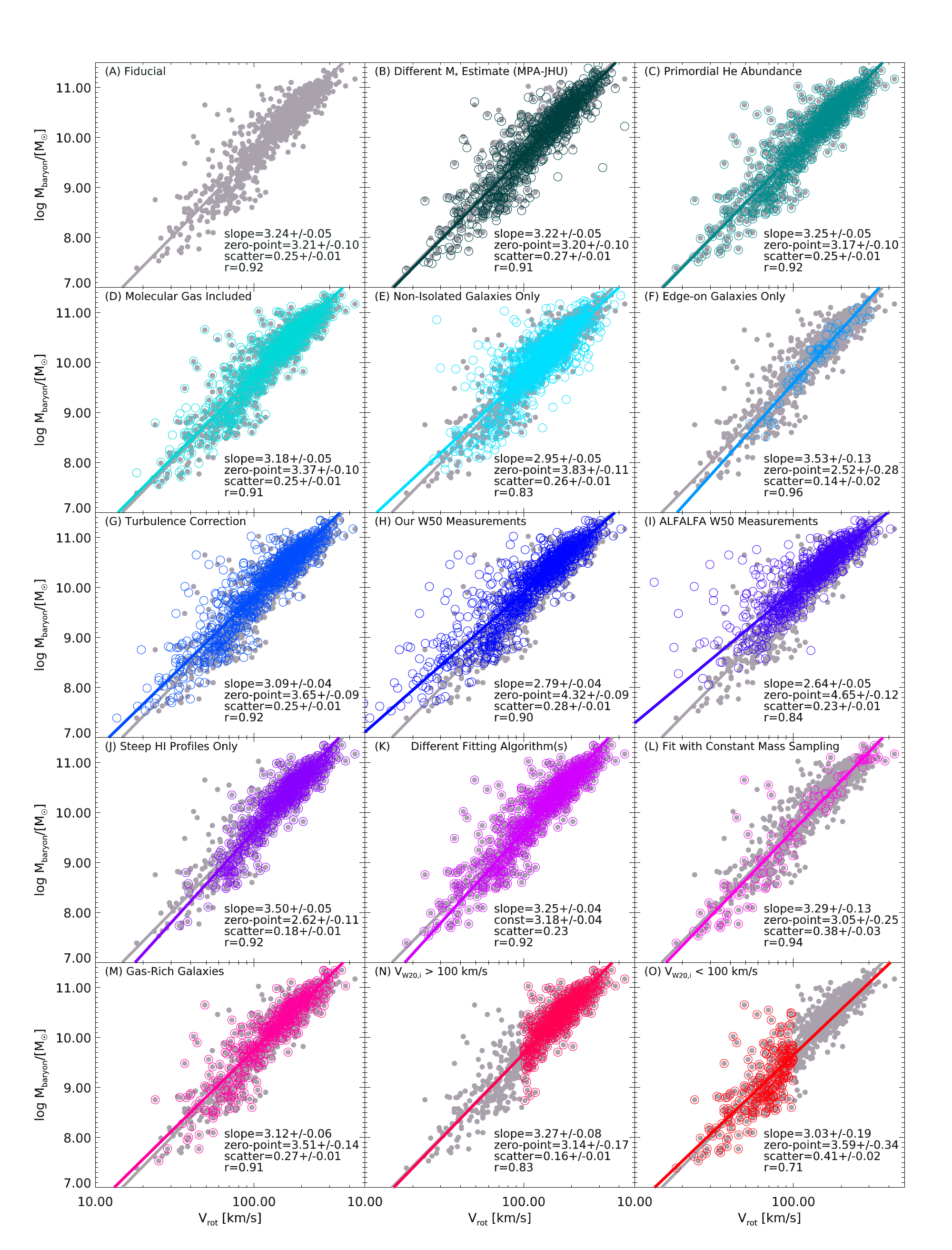}
\caption{Data and fits for each systematic in \S\ \ref{subsubsec_btf_scatslope}. Each panel is labeled according to the analysis in \S\ \ref{subsubsec_btf_scatslope}. The fiducial measurement (A) is plotted in gray in every panel for reference. Each systematic is plotted as hollow, colored dots with the fit in the corresponding color. The data used to create this figure are available.}
\label{fig_btf_syst_all}
\end{figure*}

\item {{\em{Molecular Gas Included in $M_{\rm baryon}$:}}} 
BTFR studies rarely include observations of molecular hydrogen in $M_{\rm baryon}$. Molecular gas is especially difficult to detect in low-mass galaxies due to their low CO luminosities \citep{Schruba:2012bb}. Since all of our galaxies are isolated, gas-rich, and the majority are also actively forming stars \citep{Geha:2012eu}, molecular hydrogen may contribute significantly to the gas mass of our star-forming galaxy sample. Similarly to \citet{McGaugh:2015eu}, we add an empirically derived $M_{\rm H_2}$ estimate to the total gas mass as $M_{\rm gas} = 1.4(M_{\rm HI} + M_{\rm H_2})$. We follow the method of \citet{McGaugh:2015eu} and use the SDSS observed H$\alpha$ emission star formation rates to estimate molecular gas masses as,

\begin{equation}
M_{\rm H_2} = 10^{\log_{10}{\rm SFR(H\alpha)} + 9.15} [M_{\odot}],
\label{eq_m_h2}
\end{equation}

\noindent where the total star formation rate is defined as in \citet{Martin:2001ei},

\begin{equation}
\rm{SFR(H\alpha)} = \frac{L_{\rm H\alpha}}{1.26 \times 10^{41} \rm{~ergs~s^{-1}}}10^{0.312} [M_{\odot} \rm~yr^{-1}].
\end{equation}

Since the observed H$\alpha$ flux is limited by the size of the SDSS fiber, we scale the observed H$\alpha$ flux by the ratio of $r$-band flux within the fiber to the total $r$-band flux for each galaxy. The derived SFR and $M_{H_2}$ are highly uncertain, but our goal is simply to examine how a realistic molecular gas estimate might affect the BTFR. \outstateD Including $M_{\rm H_2}$ decreases the slope slightly, but the slope and scatter remain consistent with our fiducial measurement (also see \citet{TorresFlores:2011gv}).

\item {{\em{Non-isolated Galaxies Only:}}} galaxies in dense environments (``non-isolated") are often offset from the general TFR due to processes such as triggered star formation, mass stripping, and kinematic disturbances \citep{Hinz:2003bb, MendesdeOliveira:2003bm, Cortes:2008cy, Mocz:2012en, Lelli:2015wc}. Low-mass galaxies are more susceptible to these environmental effects. In Paper~\citetalias{Bradford:2015km}, we found that including non-isolated, gas-rich galaxies does not significantly affect the slope of the BTFR. However, we found that gas-depleted, non-isolated galaxies are offset from the BTFR. Here we fit the BTFR with \nnonisogal~non-isolated galaxies \textit{only}, \nonisogallow~ of which have $V_{W20,i} < 100$\kms. \outstateE Non-isolated galaxies significantly decrease the slope of the BTFR. Interestingly, the scatter for non-isolated galaxies is only slightly larger than the fiducial measurement - implying that the sampling of the high-mass end dominates the scatter measurement (also see (O)). It is likely that these non-isolated galaxies have not yet been significantly affected by environment since we require strong \hi~detections for all galaxies.

\begin{deluxetable*}{lccccc}
\tabletypesize{\tiny}
\tablecaption{Results from Systematic Uncertainties in the BTFR Scatter and Slope}
\tablewidth{0pt}
\tablehead{
\colhead{Systematic} &
\colhead{N Galaxies} &
\colhead{Slope} &
\colhead{Zero-Point} &
\colhead{Scatter} &
\colhead{Comment}\\
\colhead{} &
\colhead{($N_{\rm Low}$/$N_{\rm High}$)} &
\colhead{} &
\colhead{} &
\colhead{} &
\colhead{Comment}\\
\colhead{(1)} &
\colhead{(2)} &
\colhead{(3)} &
\colhead{(4)} &
\colhead{(5)} &
\colhead{(6)}}
\startdata
(A) Fiducial & 930 (271$/$659) & $3.24 \pm 0.05$ & $3.21 \pm 0.10$ & $0.25 \pm 0.01$ & fiducial\\
(B) Different $M_{*}$ Estimate (MPA-JHU) & 930 (271$/$659) & $3.22 \pm 0.05$ & $3.20 \pm 0.10$ & $0.27 \pm 0.01$ & same sample as fiducial\\
(C) Primordial He Abundance & 930 (271$/$659) & $3.25 \pm 0.05$ & $3.17 \pm 0.10$ & $0.25 \pm 0.01$ & same sample as fiducial\\
(D) Molecular Gas Included & 930 (271$/$659) & $3.18 \pm 0.05$ & $3.37 \pm 0.10$ & $0.25 \pm 0.01$ & same sample as fiducial\\
(E) Non-Isolated Galaxies Only & 1308 (316$/$992) & $2.95 \pm 0.05$ & $3.83 \pm 0.11$ & $0.26 \pm 0.01$ & biased to slightly higher masses\\
(F) Edge-on Galaxies Only & 76 (17$/$59) & $3.53 \pm 0.13$ & $2.52 \pm 0.28$ & $0.14 \pm 0.02$ & biased to disky high-mass galaxies\\
(G) Turbulence Correction & 930 (323$/$607) & $3.09 \pm 0.04$ & $3.65 \pm 0.09$ & $0.25 \pm 0.01$ & same sample as fiducial\\
(H) Our W50 Measurements & 930 (345$/$585) & $2.79 \pm 0.04$ & $4.32 \pm 0.09$ & $0.28 \pm 0.01$ & same sample as fiducial\\
(I) ALFALFA W50 Measurements & 805 (214$/$591) & $2.64 \pm 0.05$ & $4.65 \pm 0.12$ & $0.23 \pm 0.01$ & similar mass distribution as fiducial\\
(J) Steep HI Profiles Only & 729 (130$/$599) & $3.50 \pm 0.05$ & $2.62 \pm 0.11$ & $0.18 \pm 0.01$ & biased to higher masses\\
(K) Different Fitting Algorithm & 930 (271$/$659) & $3.50 \pm 0.05$ & $2.62 \pm 0.11$ & $0.18 \pm 0.01$ & same sample as fiducial\\
(L) Fit with Constant Mass Sampling & 120 (68$/$52) & $3.29 \pm 0.13$ & $3.05 \pm 0.25$ & $0.38 \pm 0.03$ & sensitive to individual galaxies\\
(M) Gas-Rich Galaxies & 552 (145$/$407) & $3.12 \pm 0.06$ & $3.51 \pm 0.14$ & $0.27 \pm 0.01$ & -\\
(N) $V_{W20,i} > 100$~\kms & 659 (0$/$659) & $3.27 \pm 0.08$ & $3.14 \pm 0.17$ & $0.16 \pm 0.01$ & only high-mass galaxies\\
(O) $V_{W20,i} < 100$~\kms & 271 (271$/$0) & $3.03 \pm 0.19$ & $3.59 \pm 0.34$ & $0.41 \pm 0.02$ & only low-mass galaxies
\enddata
\tablecomments{Column definitions are: (1) Systematic ID and description, (2) total number of galaxies (number of low-mass galaxies with rotation velocity $V < 100$~\kms/number of high-mass galaxies with rotation velocity $V > 100$~\kms), (3) BTFR slope, (4) BTFR zero-point, (5) BTFR scatter, (6) comment regarding underlying distribution of galaxies.}
\label{tbl_btf_syst}
\end{deluxetable*}

\item{\em{Edge-On Galaxies Only}:} in order to examine inclination uncertainties on the BTFR, we select only edge-on galaxies from our fiducial sample. We note that choosing only edge-on galaxies severely limits the number of galaxies and biases the sample to more massive galaxies with thin disks. We impose an $b/a$ axis ratio cutoff of 0.26 to select edge-on galaxies, this value maximizes the BTFR slope and minimizes the scatter. To ensure that these galaxies are most likely edge-on and the single-dish observations probe the largest rotation velocity, we visually inspect both the \hi~spectra for double-horned profiles and the SDSS images.  We identify \nedgeon~edge-on galaxies, most of which are high-mass galaxies. The sample is severely limited to the highest S/N \hi~spectra. \outstateF This galaxy selection maximizes the BTFR slope and minimizes the scatter. The decreased scatter is most likely due to a reduction in inclination correction uncertainties. The slope is maximized because this selection biases the sample toward more disk-like galaxies, which will in turn select massive galaxies that tend to have flat rotation curves \citep{Verheijen:2001hp}. The uncertainty in the slope increases considerably because of the small sample size of this galaxy selection \citep[e.g.,][]{Sorce:2016jj}. This measurement is statistically consistent with the steep \hi~profile data sample detailed below in (J). Also see work by \citet{Papastergis:2016tl} for a detailed description of a similar process of galaxy selection that is consistent with our BTFR here before additional sample pruning to low kurtosis \hi~lines.

\item {\em{Turbulence Correction:}} non-rotational motion or ``turbulence" can increase \hi~line widths \citep{Sellwood:1999gf, Spekkens:2006iw}. This effect can be especially significant relative to $V_{\rm rot}$ at low masses \citep{Stilp:2013cv}. In \S\,\ref{subsec_btf}, we elected not to correct our rotation velocities for turbulence in our fiducial BTFR calibration. Here we examine the effects of a constant turbulent correction applied to \hi~line widths. The thermal broadening of the \hi~emission line due to thermal motions of cold and warm neutral hydrogen is expected to be 1 and 8 \kms\, respectively \citep{Tamburro:2009fy}. The typical non-rotational motion contribution to \hi~line widths is measured to be roughly 15\kms \citep[e.g.,][]{Petric:2007cs}, but this depends on resolution, mass range and observational details \citep[e.g.,][]{Zhang:2012jb, Ianjamasimanana:2015cq}.

In order to test the effect of non-circular motion on the BTFR, we use the technique typically applied in the literature to subtract so-called ``turbulent-broadening" from \hi~line widths. We use equation 6 of \citet{Trachternach:2009ff} to calculate the turbulent-corrected 20\% line widths. This equation applies a constant turbulent velocity correction of 22\kms\ to 20\% line widths \citep{Verheijen:1997vb}. At low masses, this constant term is subtracted from line widths in quadrature due to the inherently assumed Gaussian shape of the \hi~emission line. At high masses, this term is subtracted linearly from line widths since many high-mass galaxies resemble a double-peaked profile \citep{Tully:1985kb}. We note that while many low-mass galaxies do often resemble Gaussian profiles, high-resolution \hi~emission lines of low-mass galaxies are often jagged and asymmetric (see, for example, Paper~\citetalias{Bradford:2015km}, Figure 2). 

A turbulence correction has the most significant impact on the BTFR at low rotation velocities. This is due to the relative contribution of non-circular motion to the overall line widths in log space. \outstateG We find that a turbulence correction does not change the scatter of the BTFR but causes the slope to decrease by $\sim 0.15$, which is still consistent with the fiducial slope. Outliers below and to the right of the BTFR may indicate a need for turbulence corrections, but we find little motivation to apply a constant correction to our 20\% line widths.

\item {\em{Our $W_{50}$ Measurements}:} we have measured both the 20\% and 50\% \hi~line widths for our sample. 20\% line widths are always larger than 50\% line widths by about 25\kms\ in our isolated sample \citep[also see ][]{Koribalski:2004cv}. Here, we compare our fiducial BTFR measurement to one calibrated with Equation \ref{eq_v_WX}, with $X = 50$. 50\% line widths tend to push galaxies to much lower velocities than 20\% line widths because the relative change in velocity is much larger at low velocities in log space. \outstateH The effect on the slope is dramatic and results in the second smallest slope measured. The overall scatter also increases but is still consistent with the fiducial measurement.

\item {{\em{ALFALFA $W_{50}$ Measurements:}}} \hi~line widths are somewhat dependent on the method employed. Here we compare our 50\% line widths to the 50\% line widths reported by the ALFALFA survey. We use 50\% line widths because 20\% line widths are not provided in their 40\% data release catalog \citep{Haynes:2011en}. The ALFALFA BTFR slope is the smallest out of all of our fits. The ALFALFA catalog does not provide as many isolated low-mass galaxies as our data sample. \outstateI Although the change in measurement method decreases the slope significantly, this is a small change relative to the use of $W_{50}$ itself. The scatter decreases slightly due to fewer galaxies at the lowest rotation velocities.

\item {{\em{Steep \hi~Profiles Only:}}} for edge-on galaxies with both a linearly decreasing \hi~surface density and a linearly rising then flat rotation curve, each side of the \hi~emission line will be completely vertical. In contrast, the sides of an \hi~emission line of an edge-on galaxy with solid-body rotation will be parabolic. For an illustration, see \citet{Singhal:2008tp} Figures 3.4 and 3.7. Non-rotational motions (e.g., turbulence) can broaden the wings of an \hi~profile with otherwise vertical sides (see \citet{Singhal:2008tp}, Figure 3.8). By selecting galaxies from our sample with steep emission lines, we can explore the effect of inclination uncertainties, non-circular motion and rising rotation curves on the BTFR. 

To quantify the steepness of the \hi~profile, we calculate the relative difference between 50\% and 20\% rotation velocities:

\begin{equation}
\Delta V_{\rm rot} = \frac{V_{\rm W20,i} - V_{\rm W50,i}}{V_{\rm W20,i}}.
\end{equation}

\noindent Here we calibrate the BTFR with galaxies where $\Delta V_{\rm rot}$ is less than \steepcutoff, which is the cutoff where the slope is maximized while simultaneously retaining the largest number of galaxies and minimizing both the uncertainties and the scatter of the fit. Overall we retain \nsteepgalaxies~out of \NISOLATEDBTF~galaxies. Many of the BTFR outliers and most galaxies with $V_{W20,i} < 70$\kms~are removed from the sample. \outstateJ This BTFR slope is the second largest and has one of the smallest scatters of all of our measurements. This result is interesting as this slope is similar to the edge-on value in (F), which are both closest to values in the literature that use $V_{\rm flat}$ determined from resolved rotation curves. This may be a way to pre-select galaxies for resolved \hi~synthesis surveys with extended rotation curves. However, it is known that line widths are in fact not a perfect determination of $V_{\rm flat}$ even when $V_{\rm flat}$ is observable \citep{McGaugh:2012ev}.
 
\item {{\em{Different Fitting Algorithm:}}} for every fit in this paper, we use the Bayesian linear regression algorithm of \citet{Kelly:2007bv}, as described in \S\ \ref{subsec_fit_lin}. \citet{Weiner:2006bv} have explored fitting algorithms for the TFR, which is also enlightening with respect to the BTFR. They find $\chi^2$ minimization \citep[e.g.,][]{Press:1992vz}, which does not model scatter, increases the TFR slope as scatter increases - while generalized least squares, maximum likelihood and bivariate correlated errors with scatter (BCES) are all relatively successful at recovering a known relation. These authors also find that bisector fits \citep[e.g.,][]{Isobe:1990ft}, which measure the bisecting line of the forward and inverse relations, can be severely affected by selection biases. Indeed, forward and reverse fitting algorithms can be heavily influenced by Malmquist bias \citep{Strauss:1995dj, Willick:1997gg}. 

We first compare our fiducial fit to the fitting algorithms implemented in \texttt{SIXLIN} \citep{Isobe:1990ft}. These fitting algorithms are commonly used in fitting galaxy scaling relations like the BTFR \citep[e.g.,][]{Courteau:2007ek, Hall:2012hh, Zaritsky:2014fr} and are also used to compare BTFR fits against other modern algorithms \citep[e.g.,][]{McGaugh:2012ev}. The \texttt{SIXLIN} algorithm fits forward ordinary least squares, inverse ordinary least squares, ordinary least squares bisector, orthogonal, reduced major axis and mean ordinary least squares to a linear model with each fit weighted by the uncertainty in $M_{\rm baryon}$. For each algorithm, we measure slopes of \slfwdolsslope, \slrevolsslope, \slbisolsslope, \slortholsslope, \slredmajslope~and \slmeanolsslope, respectively. The method used to fit the BTFR can change the slope by $\pm$\sldiffslope. Since observed scatter in the BTFR are not included in these fitting algorithms, we plot these fits in Figure \ref{fig_btf_syst} as downward-pointing triangles on the top x-axis. The slopes of the bisector, reduced major axis and mean ordinary least squares fits are all consistent with the \citet{Kelly:2007bv} algorithm, while the forward ordinary least squares, inverse ordinary least squares and orthogonal fits are not. We note that we obtain similar results when fitting the edge-on galaxies in (F). Because these fits do not include uncertainty in both the x- and y-axis and because they do not fit for intrinsic scatter, we elect not to report these fits as official results of our study. We retain the slopes in Figure \ref{fig_btf_syst} for illustrative purposes only.

We also compare our fiducial results using the fitting wrapper \texttt{mpfitxy} by \citet{Williams:2010hn} (see their \S\ 4) which is built around the Levenberg--Marquardt least squares fitting routine \texttt{mpfit} by \citet{Markwardt:2009wq}. This fitting algorithm takes into account the errors in both variables and also models scatter in the BTFR. This algorithm was coded by \citet{Williams:2010hn} to compare the TFR of S0 galaxies to spiral galaxies. \outstateK This algorithm is consistent with our fiducial model. Even though the \texttt{mpfitxy} fitting algorithm is consistent with our fiducial model, several of the results from \texttt{SIXLIN} are not. The fitting algorithm is therefore a significant source of uncertainty in measuring the BTFR slope and the algorithms in \texttt{SIXLIN} are indeed applied in the literature. Fitting algorithm is therefore an additional complication when comparing BTFR fits.

\item {{\em{Fit with Constant Mass Sampling:}}} while our \hi~observations have significantly increased the number of isolated low-mass galaxies with single-dish \hi~observations (Paper I), our data remains biased toward the high-mass end of the BTFR due to the flux-limited nature of the ALFALFA survey. We test the effect of uneven sampling and the effectiveness of our linear fitting algorithm by uniformly and randomly sampling the data set.

We generate a data set with constant mass sampling by randomly selecting galaxies from uniformly distributed mass bins. The mass bin size is defined as the width of the lowest mass bin that includes 20 galaxies. We then randomly select 20 galaxies from each subsequent mass bin of the same size (0.72 dex per bin) but at greater $M_{\rm baryon}$. \outstateL The scatter increases because it is dominated by the low-mass end. The scatter measurement is also highly dependent on the result of the random draw. The slope increases but is highly uncertain and consistent with our fiducial measurement. This implies the slope derived from the \citet{Kelly:2007bv} algorithm is robust to our sampling of galaxies. We also note this result is consistent with simulation as in \citet{Sorce:2016jj}.

\item {\em{Gas-Rich Galaxies}}: resolved \hi~observations are biased toward galaxies that are \hi-rich. It may be that galaxies at fixed rotation velocity have different fractions of their baryons in hot ionized or cold molecular gas relative to the atomic gas than we have measured. Also, despite the gas fraction of the galaxies in our sample, we may still be affected by stellar mass estimates. Therefore we are interested in the effect of atomic gas fraction on the BTFR slope and scatter. We select galaxies with observed gas masses greater than the gas mass predicted from the relation between stellar mass and atomic gas mass in Paper~\citetalias{Bradford:2015km}, table 3. For galaxies with $M_* < 10^{8.6} M_{\odot}$, $M_{\rm gas} > (1.052 \log{M_*} + 0.236)$, for galaxies with $M_* \geq 10^{8.6} M_{\odot}$, $M_{\rm gas} > (0.461 \log{M_*} + 5.329)$. \outstateM Selecting only gas-rich galaxies slightly increases the scatter and decreases the slope. This gas-rich galaxy sample is consistent with the fiducial fit. This result also supports the conclusion of (B).

\item {{\em{$V_{W20,i} > 100$\kms:}}} several of the results above suggest that low-mass galaxies drive the scatter in the BTFR. Here we examine differences in the BTFR for galaxies with large rotational velocities compared to galaxies with small rotational velocities. We divide our sample on $V_{\rm rot}$ because we are interested in the scatter in $M_{\rm baryon}$ as a function of $V_{\rm rot}$ and cutting on $M_{\rm baryon}$ will affect the measured scatter. We choose galaxies with $V_{W20,i} > 100$\kms as our ``high-mass" sample. We have \NHMISOLATEDBTF~isolated galaxies at high masses, accounting for \HIGHMASSPERC \% of our total sample. \outstateN High-mass galaxies produce the same slope compared to our fiducial measurement but the scatter decreases significantly to 0.16. As confirmed below, a large fraction of the scatter in the fiducial relationship must therefore driven by galaxies with $V_{W20,i} < 100$\kms.

\item {{\em{$V_{W20,i} < 100$\kms:}}} we choose galaxies with $V_{W20,i} < 100$\kms~as our ``low-mass" sample. We have \NLMISOLATEDBTF~isolated galaxies at low masses, accounting for \LOWMASSPERC \% of our total sample. Low-mass galaxies tend to have either rising rotation curves at the outermost radius or uncertain maximum and asymptotic velocities which can dramatically affect the slope and scatter of the BTFR \citep[e.g.,][]{Verheijen:2001hp}. \outstateO For galaxies with $V_{W20,i} < 100$\kms, we measure a smaller slope that is highly uncertain but statistically consistent with our fiducial measurement. The scatter roughly doubles and is inconsistent with our fiducial measurement. This is the only measurement where a significant increase in the scatter occurs and also results in the least correlated relation (smallest Pearson's r.) This result confirms that most of the scatter in our fiducial measurement comes from galaxies with $V_{W20,i} < 100$\kms (also see \citet{Gurovich:2010ha}).

\end{enumerate}

\begin{deluxetable*}{cccccc}
\tabletypesize{\tiny}
\tablecaption{BTFR Fits from the Literature in Order of Rotation Velocity Definition}
\tablewidth{0pt}
\tablehead{
\colhead{Abbreviation} &
\colhead{Reference} &
\colhead{Slope*} &
\colhead{Rotational Velocity} &
\colhead{Baryonic Mass Range**} &
\colhead{Velocity Tracer}\\
\colhead{ } &
\colhead{ } &
\colhead{ } &
\colhead{} &
\colhead{$\log{M_{\odot}}$} &
\colhead{ }\\
\colhead{(1)} &
\colhead{(2)} &
\colhead{(3)} &
\colhead{(4)} &
\colhead{(5)} &
\colhead{(6)}}
\startdata
N07 & \citet{Noordermeer:2007cc} & $3.04 \pm 0.08$ &$V_{W20}$ & 9.5 to 11.5 & \hi~\\
AR08 & \citet{AvilaReese:2008gz} & $3.27 \pm 0.13$ &$V_{W20}$ & 9.1 to 11.6 & \hi~\\
G10 & \citet{Gurovich:2010ha} & $3.20 \pm 0.10$ &$V_{W20}$ & 7.5 to 11.4 & \hi~\\
M12 & \citet{McGaugh:2012ev} & $3.41 \pm 0.08$ &$V_{W20}$ & 6.4 to 11.3 & \hi~\\
\hline
M08 & \citet{Meyer:2008fe} & $3.91 \pm 0.13$ &$V_{W50}$ & 9.1 to 11.1 & \hi~\\
C12 & \citet{Catinella:2012ea} & $4.22 \pm 0.10$ &$V_{W50}$ & 10.1 to 11.4 & \hi~\\
H12 & \citet{Hall:2012hh} & $3.41 \pm 0.10$ &$V_{W50}$ & 8.0 to 11.3 & \hi~\\
Z14 & \citet{Zaritsky:2014fr} & $3.50 \pm 0.20$ &$V_{W50}$ & 8.6 to 11.8 & \hi~\\
P16 & \citet{Papastergis:2016tl} & $3.94 \pm 0.14$ &$V_{W50}$ & 8.3 to 10.6 & \hi~\\
\hline
M00 & \citet{McGaugh:2000hx} & $3.98 \pm 0.12$ &Mixed & 6.7 to 11.5 & \hi~\\
G06 & \citet{Geha:2006jx} & $3.70 \pm 0.14$ &Mixed & 8.0 to 11.2 & \hi~\\
D07 & \citet{DeRijcke:2007bp} & $3.15 \pm 0.07$ &Mixed & 7.6 to 11.7 & Stars and \hi~\\
\hline
B01 & \citet{Bell:2001hv} & $3.53 \pm 2.80$ &$V_{\rm flat}$ & 9.4 to 11.0 & \hi~\\
V01 & \citet{Verheijen:2001hp} & $4.00$ &$V_{\rm flat}$ & 9.8 to 11.4 & \hi~\\
M05 & \citet{McGaugh:2005bc} & $4.00 \pm 0.15$ &$V_{\rm flat}$ & 8.6 to 11.6 & \hi~\\
N07 & \citet{Noordermeer:2007cc} & $3.38 \pm 0.10$ &$V_{\rm flat}$ & 9.5 to 11.5 & \hi~and emission lines \\
S09 & \citet{Stark:2009ks} & $3.94 \pm 0.10$ &$V_{\rm flat}$ & 7.9 to 11.5 & mixed but mostly \hi~\\
M12 & \citet{McGaugh:2012ev} & $3.94 \pm 0.10$ &$V_{\rm flat}$ & 6.4 to 11.3 & \hi~\\
M15 & \citet{McGaugh:2015eu} & $4.04 \pm 0.09$ &$V_{\rm flat}$ & 6.0 to 11.5 & \hi~\\
B15 & \citet{Bottema:2015ji} & $3.70 \pm 0.20$ &$V_{\rm flat}$ & 8.6 to 11.2 & \hi~\\
\hline
K04 & \citet{Kregel:2005fx} & $3.23 \pm 0.36$ &$V_{\rm max}$ & 9.3 to 11.7 & \hi~\\
K06 & \citet{Kassin:2006ce} & $3.40 \pm 0.30$ &$V_{\rm max}$ & 9.9 to 11.6 & \hi~and H$\alpha$ \\
N07 & \citet{Noordermeer:2007cc} & $3.05 \pm 0.09$ &$V_{\rm max}$ & 9.5 to 11.5 & \hi~and emission lines \\
TF11 & \citet{TorresFlores:2011gv} & $3.64 \pm 0.28$ &$V_{\rm max}$ & 8.9 to 11.4 & H-alpha \\
B15 & \citet{Bottema:2015ji} & $4.30 \pm 0.40$ &$V_{\rm max}$ & 8.6 to 11.2 & HI
\enddata
\tablecomments{Column definitions are: (1) abbreviation used in the literature figure, (2) literature reference, (3) baryonic Tully--Fisher slope with uncertainties, (4) baryonic mass range over which the relation was measured, and (5) rotational velocity tracer. *Slopes are often reported using multiple methods in each study, we attempted to select slopes that are most similar to our fiducial model. Refer to the original papers for details. Several slopes are reported for the inverse relation V--M which we invert. We assume a flat uncertainty of 0.1 in the slope if the uncertainty is not reported. **Baryonic masses ranges are often estimated by eye these ranges are to serve as a rough guide.}
\label{tbl_lit}
\end{deluxetable*}

Varying the input assumptions and sample selection within our single, homogeneously measured data set, we obtain slopes ranging from \minsystslope~to \maxsystslope~and scatter ranging from \minsystscatter~to \maxsystscatter. The most significant systematic is the definition of rotation velocity, as in (H) and (I). The details of the fitting algorithm in (K) can also affect the slope measurement considerably, particularly forward, inverse or orthogonal ordinary least square fitting where scatter is not modeled and errors are not included in both dimensions. Without changing the $M_{\rm baryon}$ definition, the fitting algorithm or the $V_{\rm rot}$ definition from the fiducial sample, the greatest difference in slope occurs if we only include non-isolated galaxies in the sample as in (E) or if we only include galaxies with steep \hi~emission profiles as in (J) or edge-on galaxies as in (F). A small sample size can also increase the uncertainties in the BTFR measurements as in (F), (L) and (O).

It is apparent from these results that low-mass galaxies drive the scatter and the uncertainty in the slope of the BTFR \citep[also see][]{Saintonge:2010gg}. While galaxies with $V_{\rm rot} < 100$\kms~are consistent with the BTFR of galaxies with larger $V_{\rm rot}$, these galaxies significantly increase the scatter in the BTFR regardless of input assumptions. While the scatter we measure at low masses may be a natural consequence of galaxy formation \citep{Dutton:2007he, Dutton:2009dn}, it may also be due to underestimated observational uncertainties, the selection of galaxies we use to calibrate the BTFR and the underlying rotation curves of low-mass galaxies. For example, while the random uncertainties of b/a axis ratios are insignificant, the systematic uncertainties are unknown and may be significant at low masses where morphologies are less disk-like \citep{Sancisi:1976ts, Papastergis:2011ie, Obreschkow:2013ho}. The uncertainty in the derived inclinations may contribute significantly to the subsequent uncertainty in $V_{W20,i}$. Underestimating the uncertainty in $V_{W20,i}$ may therefore drive the low-mass scatter in the BTFR. This is supported by the edge-on galaxy sample we select in (F) where only several low-mass galaxies are classified as edge-on.

Various combinations of the systematics (A) through (O) above, can result in even more dramatic effects on the BTFR. Exploring all of the various combinations of these systematics is beyond the scope of this paper. We simply illustrate here that rotation velocity definition, sample selection and fitting algorithm can dramatically affect the observed BTFR as calibrated with \hi~line widths. While we measure a small uncertainty in the slope of our fiducial BTFR of 0.05, a systematic uncertainty of 0.25 would be more realistic. This systematic uncertainty would be more in line with the values of (E) to (F) above. We next compare our results to the literature in an attempt to check the consistency of our results.

\subsection{Comparison to the Literature}
\label{subsec_lit}

\begin{figure*}[t!]
\epsscale{1.19}
\plotone{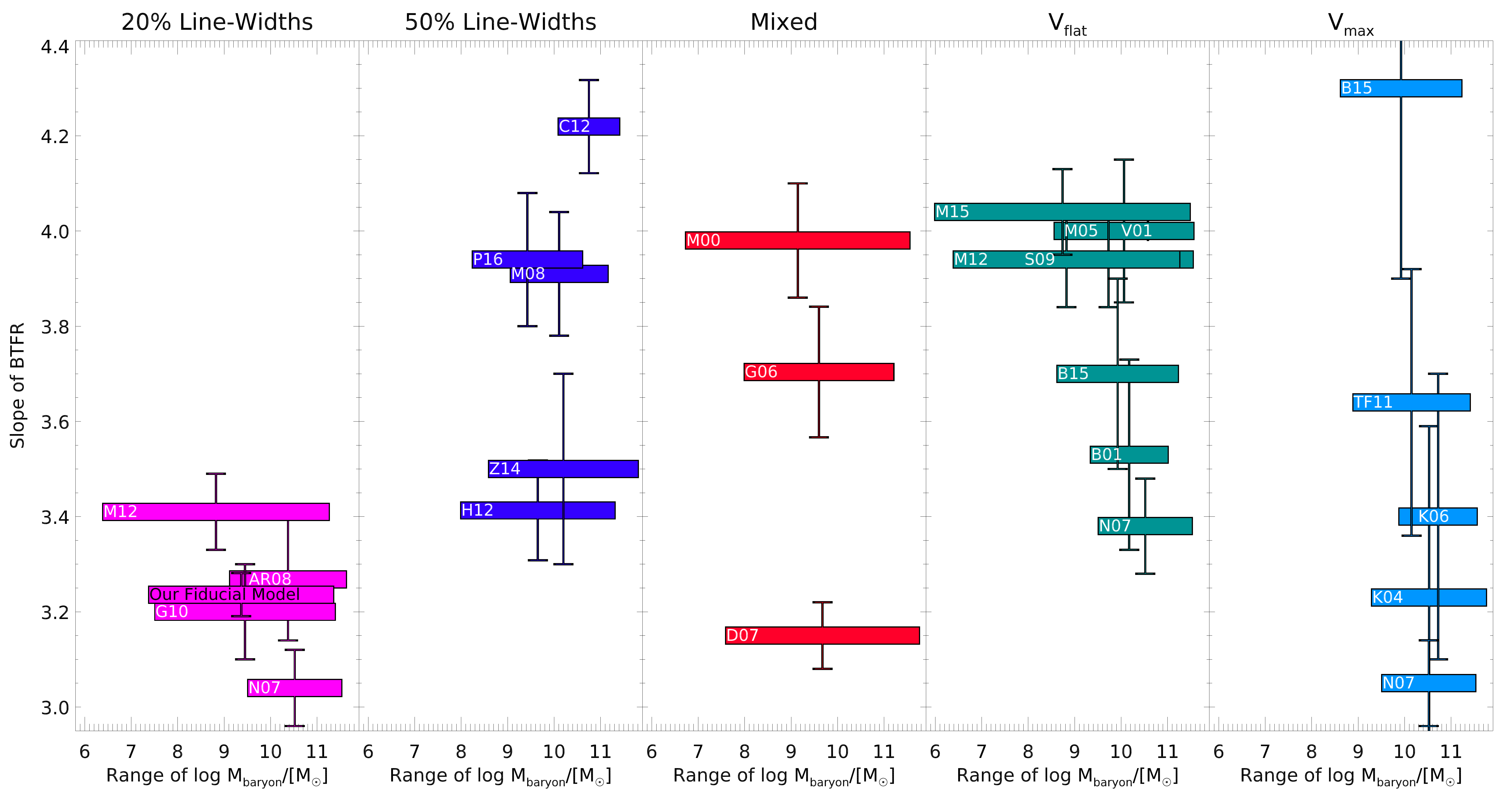}
\caption{Decades of $M_{\rm baryon}$ used to make the BTFR fit versus slope from the literature as detailed in Table \ref{tbl_lit}. Each horizontal bar represents a different measurement of the BTFR. Each measurement is not necessarily independent and may contain a significant fraction of galaxies as another measurement. Error bars are the reported uncertainty in the slope. Each color and panel represents a different rotation velocity measurement, with pink representing 20\% \hi~line widths, dark blue representing 50\% \hi~line widths, red representing mixed measurements, aqua representing the asymptotic velocity $V_{\rm flat}$ and light blue representing the maximum rotation velocity $V_{\rm max}$. The two left-most panels present the BTFRs measured using only unresolved \hi~line widths. The right-most panels represent BTFRs measured with resolved rotation curves either with single-slit nebular emission lines or \hi~interferometry. Our fiducial model is also shown for comparison. The data used to create this figure are available.}
\label{fig_lit_data}
\end{figure*}

BTRF slopes are typically measured using a variety of rotation velocity measurements, mass ranges, stellar mass definitions, and fitting methods. Rotation velocities are derived from \hi~line widths \citep[e.g.,][]{Gurovich:2010ha}, single-slit nebular emission line rotation curves \citep[e.g.,][]{Kassin:2006ce}, resolved \hi~emission rotation curves \citep[e.g.,][]{Verheijen:2001hp},  or some combination of these different rotation velocity measurements \citep[e.g.,][]{Geha:2006jx}. The $M_{\rm baryon}$ range over which the BTFR is measured varies from 1.3 dex \citep{Catinella:2012ea} to 5.5 dex \citep{McGaugh:2015eu} and tends to cover 3 dex in most studies with samples that are heavily biased toward the massive end of the relation. Many studies use either non-weighted $\chi^2$ \citep[e.g.,][]{Noordermeer:2007cc}, bisector least squares \citep[e.g.,][]{Kassin:2006ce}, inverse least squares \citep{Catinella:2012ea}, bivariate least squares \citep{Meyer:2008fe} or a variety of other astronomical fitting algorithms \citep[e.g.,][]{Tremaine:2002hd, Weiner:2006bv, Kelly:2007bv, Saintonge:2010gg, Hall:2012hh, Cappellari:2013jz}. 

Despite these variations, observational calibrations, and theoretical predictions of the BTFR are often compared either directly to BTFR data in the literature or, more commonly, to the derived slope, zero-point and scatter of literature BTFR fits - regardless of measurement methods \citep[e.g.,][]{Aumer:2013fm, Kang:2013dc, Vogelsberger:2014tw, Zaritsky:2014fr, denHeijer:2015bq}. The disparity in such comparisons is usually due to a subtle difference in $V_{\rm rot}$ definition. Low-mass galaxies are especially susceptible to definitions of $V_{\rm rot}$ due to their low \hi~column densities and typically rising rotation curves in \hi~\citep{Brook:2016gh}.

We explore the importance of the $V_{\rm rot}$ definition and baryonic mass range on BTFR measurements in the literature. Since many literature studies do not report scatter in the BTFR, the zero-point of the BTFR or the details of their fitting algorithm, we focus here only on the slope. We summarize the results of \nliteraturemeasures~BTFR studies in Figure \ref{fig_lit_data}, Table~\ref{tbl_lit} and the discussion below.  \citet{TorresFlores:2011gv} also provide a comprehensive list of fits to the BTFR from the literature (see their Table 3). 

We divide the relevant literature into unresolved, mixed, and resolved rotational velocities where unresolved measurements are typically derived from single-dish \hi~line widths, resolved velocities tend to be derived from \hi~rotation curves and mixed studies use both unresolved and resolved rotational velocities to cover a larger mass range than any single homogeneous sample. The slopes we report below are often selected from several different calibrations of the BTFR that have been made within each study. For most of the literature slopes below, we report the value with the least scatter or the calibration that is reported in each study's conclusion. When possible, we select slopes from each study based on similarities between rotation velocity definitions and similar methods as our fiducial measurement in order to disentangle the differences between measurements and to minimize any supposed controversy. In some cases we have inverted the measured slope if necessary.

\subsubsection{Unresolved line widths} 

line widths, especially at small rotation velocities, are affected by measurement method, non-rotational motion, S/N, underlying galaxy rotation curve shapes, and baryonic tracer density. line width measurement methods can affect subsequent rotation velocity measurements. It is common to measure the \hi~line width as some fraction of the \hi~flux density \citep[e.g.,][Paper I]{Koribalski:2004cv} or a fraction of the average of multiple peaks or ``profile horns" \citep{Haynes:1999gl, Springob:2005db}.  The precise meaning of \hi~line widths vary from study to study \citep[e.g.,][]{Courteau:1997ic, Verheijen:1997vb, Verheijen:2001gr, Blanton:2008il, Courtois:2009ix, Zavala:2009dh}. It is unclear if \hi~line widths (20\% or 50\%) consistently correlate with a characteristic resolved rotation velocity (e.g., $V_{\rm flat}$, $V_{\rm max}$), especially at the low- and high-mass extremes of the BTFR \citep{Verheijen:2001gr, Noordermeer:2007cc, Brook:2016gh}. Therefore we would expect some variation in BTFR measurements depending on the details of the \hi~line width definition and the S/N of the low-mass end of the BTFR sample.

Here we examine BTFR measurements made with unresolved line widths in the left two panels of Figure \ref{fig_lit_data}. Despite the uncertainties we have discussed above, our fiducial model is consistent with the 20\% line width measurements of \citet{Gurovich:2010ha}, \citet{AvilaReese:2008gz} and \citet{McGaugh:2012ev}. We plot the inverse of the \citet{AvilaReese:2008gz} slope where they have employed an orthogonal fitting algorithm and mostly 20\% line widths. We note that their inverse slopes range from 0.303 (forward fitting) to 0.333 (inverse fitting), depending on the fitting algorithm. \citet{Gurovich:2010ha} measure the 20\% line width and employ the \citet{Press:1992vz} FITXY algorithm. The \citet{Noordermeer:2007cc} measurement is slightly smaller than other 20\% line width studies, but is within $2\sigma$ of our fiducial measurement. The \citet{Noordermeer:2007cc} measurement is interesting because their BTFR calibration to the high-mass end produces similar slopes using their 20\% line widths and $V_{\rm max}$.

Curiously, the 50\% width measurements (in blue) tend to produce larger slopes than the 20\% measurements. Since 50\% widths are always smaller than 20\% widths, we would expect (and have measured in \S\ \ref{subsubsec_btf_scatslope}) the exact opposite trend. \citet{Zaritsky:2014fr} and \citet{Hall:2012hh} are both within $2\sigma$ of our fiducial measurement and use different methods for calculating the 50\% line width (see \citet{Courtois:2011kva} and \citet{Springob:2005db} methods, respectively). \citet{Hall:2012hh} employ an orthogonal fitting algorithm to the inverse relation and 50\% line widths. We match both the slope and zero-point within 1$\sigma$ using the same measurement method as \citet{Hall:2012hh} but applied to our data. \citet{Zaritsky:2014fr} examine several systematics in their BTFR and then perform least ordinary least squares fitting to obtain slopes between 3.3 and 3.7 with a final quoted slope of $3.5\pm0.2$. \citet{Zaritsky:2014fr} suggests that the difference between their study and resolved BTFRs is due to stellar mass estimates.

The remaining three 50\% line width BTFRs are inconsistent with our fiducial measurement. The \citet{Papastergis:2016tl} value has been pruned to edge-on galaxies similarly to our systematic (F) but with low kurtosis \hi~lines. These authors believe that pruning the ALFALFA sample allows them to select galaxies where the 50\% line widths effectively measure $V_{\rm flat}$. The discrepancies with the other two data points may be due to the baryonic mass regime over which the slopes were measured and/or fitting algorithm used. For example, the \citet{Catinella:2012ea} sample is restricted to galaxies with $M_* > 10^{10} M_{\odot}$ and they use an inverse least squares fit to the inverse of the BTFR we measure. If we compare our galaxy sample for the same mass range, nearly all of our data set is consistent with their data set when 20\% line widths are used. However, we are unable to precisely replicate their fit because the \citet{Catinella:2012ea} sample contains several galaxies at high-masses and lower $V_{\rm rot}$ than our sample, and our sample contains several galaxies at high-masses and higher $V_{\rm rot}$ than their sample. This demonstrates the BTFR sensitivity to sample selection, velocity measurement and baryonic mass range. \citet{Meyer:2008fe} report two slopes of 4.35 in the K-band and 3.91 in the B-band, demonstrating that the BTFR slope is extremely sensitive to stellar mass estimates for high-mass galaxy samples (also see \citet{McGaugh:2005bc}). Our goal is not to evaluate the precise reason for the discrepancy but to point out that BTFRs derived from line widths are heavily influenced by the details and nuances of the study. The literature slope measurements that are consistent within $2\sigma$ of our fiducial model range from $3.04\pm0.08$ to $3.5\pm0.2$. Interestingly, this range corresponds to the same range of slopes we measure in \S\ \ref{subsubsec_btf_scatslope} using 20\% line widths between (E) and (F).

\subsubsection{Resolved Rotation Curves} 
\label{subsubsec_res_lit}

Resolved BTFR studies are restricted to relatively gas-rich galaxies with measurable rotation curves where reported velocities are either measured at a radius where the rotation curves asymptote ($V_{\rm flat}$), where the rotation curves maximize ($V_{\rm max}$) or simply the last measured data point in the rotation curve ($V_{\rm last}$). The particular velocity definition may impose a selection effect on the data sample used to calibrate the BTFR \citep{Verheijen:2001hp, Gurovich:2010ha}. In other words, $V_{\rm max}$ or $V_{\rm last}$ may shift low-mass galaxies to lower rotational velocities and high-mass galaxies to higher rotational velocities - the effect of which would also be reflected in the unresolved \hi~line widths discussed above \citep{Verheijen:2001hp}. If the BTFR is \textit{strictly defined} as the relationship between the total gravitational potential and the observed $M_{\rm baryon}$ of spiral galaxies, then resolved rotation curves should produce the most accurate BTFR. 

Resolved studies that measure $V_{\rm flat}$ from \hi~rotation curves and stellar masses based on infrared luminosities tend to produce slopes near 4 with little to no scatter. If galaxy samples are not carefully selected for asymptotic rotation curves, the BTFR will be affected by galaxies with rising or falling rotation curves. This becomes a challenging observational endeavor at low masses due to the intrinsic low \hi~surface densities of low-mass galaxies. Flat rotation curves have variable definitions in the literature, but are generally identified when the last few measurable rotation velocities at large radii only differ by a few percent. In many low-mass galaxies, neither $V_{\rm flat}$ nor $V_{\rm max}$ is measurable because rotation curves of low-mass galaxies are often rising at the last measured radius, \citep[e.g.,][Figure 1]{Stark:2009ks}. If the flat part of a galaxy's rotation curve is measurable, then we might expect any rotation velocity measurement ($V_{\rm max}$, $V_{\rm flat}$, $V_{W20,i}$ and $V_{W50,i}$) to produce roughly the same BTFR slope as in \citet{Courteau:1997ic} and \citet{Trachternach:2009ff} (see their Figure 8). However, this is not always a guarantee \citep{McGaugh:2012ev}.

In the right panels of Figure~\ref{fig_lit_data}, the reported slopes for either $V_{\rm max}$ or $V_{\rm flat}$ are as varied as the unresolved studies. Aside from \citet{Bottema:2015ji}, $V_{\rm max}$ tends to produce shallower slopes that are similar to line width slopes. \citet{Noordermeer:2007cc} report slopes using both estimators and find the BTFR slope using $V_{\rm max}$ is 0.33 smaller than the one made with $V_{\rm flat}$. Interestingly, this is similar to the difference between our fiducial slope and the slope we measure for steep \hi~profiles and edge-on galaxies in the previous section. Given the published uncertainties added in quadrature to ours, \citet{Bell:2001hv}, \citet{Kregel:2005fx}, \citet{Kassin:2006ce}, \citet{Noordermeer:2007cc}, \citet{TorresFlores:2011gv} and \citet{Bottema:2015ji} are all within $3\sigma$ of our fiducial relation. \citet{Verheijen:2001hp}, \citet{McGaugh:2005bc}, \citet{Stark:2009ks}, \citet{McGaugh:2012ev} and \citet{McGaugh:2015eu} are all statistically inconsistent with our fiducial measurement, most likely because these studies only include galaxies with well-measured $V_{\rm flat}$. We note that the \citet{McGaugh:2015eu} relation we chose is from their table 5, which includes data from \citet{McGaugh:2012ev}. These authors make a separate calibration that is consistent with this measurement but without \citet{McGaugh:2012ev} data.

We note that many of the resolved studies in Figure~\ref{fig_lit_data} use similar data sets and analysis techniques. For example, nearly 80\% of the galaxies in the \citet{Stark:2009ks} measurement also appear in \citet{McGaugh:2000hx}, \citet{McGaugh:2005bc} and \citet{McGaugh:2012ev}. Overall, \citet{McGaugh:2000hx, Verheijen:2001hp, McGaugh:2005bc, Stark:2009ks, McGaugh:2012ev, McGaugh:2015eu} and \citet{Bottema:2015ji} fit the BTFR to some subset of roughly 200 galaxies (see \citet{ Lelli:2016ed} for the most recent list of data sources). Aside from \citet{Trachternach:2009ff}, which overlaps almost entirely with \citet{Stark:2009ks}, we plot nearly all resolved studies of the BTFR that we are aware of, regardless of the number of overlapping galaxies. This is not meant to be a criticism of this technique, but simply to illustrate that these measurements are not completely independent and that it is hard to measure $V_{\rm flat}$ using resolved \hi~rotation curves. In truth, these may be the only ``correct" measurements of the BTFR as defined at the beginning of this section.

In contrast to these resolved studies, the unresolved 20\% line width measurements partly overlap in \citet{AvilaReese:2008gz} and \citet{Noordermeer:2007cc}, since they measure the BTFR from a similar pool of galaxies as the $V_{\rm flat}$ measurements, but \citet{Gurovich:2010ha} use data from the southern-sky HIPASS survey and our calibration is derived from a combination of the northern-sky ALFALFA survey and our own data. Therefore the 20\% line width measurements here provide at minimum three relatively independent calibrations of the BTFR. This simply illustrates that point that unresolved studies offer a greater number of galaxies to compare to predictions of galaxy formation.

\subsubsection{Mixed Studies} 
Several studies have mixed unresolved line widths and resolved rotation curves in order to calibrate the BTFR over the largest possible baryonic mass range (see Figure~\ref{fig_lit_data}, middle panel). These are the most difficult measurements to reproduce theoretically and are all inconsistent with one another at the 1$\sigma$ level (middle panel, Figure~\ref{fig_lit_data}). We do not recommend using mixed line width fits to compare to simulations, nor do we recommend this practice in calibrating the BTFR. For example, \citet{Geha:2006jx} mix 20\% \hi~line widths of low-mass galaxies that have linearly subtracted turbulence corrections with resolved high-mass galaxy rotation curves and single-dish observations from the literature. \citet{Geha:2006jx} measure an inverted slope of 1.9 using just their low-mass sample, but they measure an inverted slope of 3.7 when higher mass galaxies are folded in.

\section{Discussion}
\label{sec_discussion}

We have explored the BTFR using a homogeneous catalog of \NISOLATEDBTF~isolated galaxies. We fit the line width BTFR using inclination-corrected 20\% \hi~line widths and the \citet{Kelly:2007bv} fitting algorithm over a baryonic mass range $10^{\minbtfbaryonicmass} < M_{\rm baryon} < 10^{\maxbtfbaryonicmass} M_{\odot}$. We examine the impact of rotation velocity definition, $M_{\rm baryon}$ definition, sample selection, environment and fitting algorithm on the line width BTFR fit. We also perform a comparison to \nliteraturemeasures~BTFR studies, including relations measured using both resolved and unresolved rotation velocities. The results of this work are as follows:

\begin{enumerate}
\item{We measure a fiducial BTFR slope of $\btfslope~\pm~\btfslopeerr$, zero-point of $\btfconst \pm \btfconsterr$, observed scatter of $\btfscat \pm \btfscaterr$. These uncertainties are random and do not include systematic uncertainties.}
\item{We obtain slopes between \minsystslope~and \maxsystslope~within our data set, depending mostly on the rotation velocity definition but also on the galaxy sample selection and the linear fitting algorithm.}
\item{We measure scatter between \minsystscatter~and \maxsystscatter~within our data set. Galaxies with {$V_{W20,i} < 100$\kms} drive most of the observed scatter in the BTFR. This increase is most likely due to underestimated inclination uncertainties.}
\item{In the literature, the choice of rotation velocity measurement has a large effect on the reported BTFR slope. 20\% \hi~line widths produce a median slope of \wtwentymedslope, 50\% line widths produce a median slope of \wfiftymedslope, maximum velocities produce a median slope of \maxmedslope, and flattening velocities produce a median slope of \flatmedslope. These slopes are also influenced by sample selection, particularly the baryonic mass range over which the relation is measured.}
\item{While we measure a random uncertainty in the slope of our fiducial slope measurement of \btfslopeerr, we can derive a systematic uncertainty of 0.25. This systematic uncertainty is consistent with the range of slopes produced by our fiducial data set given the various assumptions and calibrations studies above. This systematic uncertainty brings our fiducial measurement into agreement with nearly all line width BTFR studies.}
\item{When comparing observed or simulated BTFRs, we suggest measuring the relations over the same mass range with the same fitting algorithm, sample selection, and rotation velocity definition.}
\end{enumerate}

The difference in the literature slopes derived from 20\% line widths and $V_{\rm flat}$ are most likely due to a systematic shift in low-mass galaxies toward lower rotational velocities in the unresolved samples. If the gas disks of low-mass galaxies tend to be truncated relative to more massive galaxies, this would mean that unresolved \hi~measurements may systematically underestimate rotation velocities of low-mass galaxies \citep{McGaugh:2012ev, Brook:2016gh}. The effect would be to shift low-mass galaxies off of the high-mass relation to lower circular velocities and in turn decrease the BTFR slope \citep{Verheijen:2001hp}. We avoid some of these effects in \S\ \ref{subsubsec_btf_scatslope} (F) and (J) by limiting our sample to edge-on galaxies with double-horned \hi~profiles and by selecting galaxies with very steep \hi~profiles; which produce the largest slopes and smallest scatter measurements. This comes at a cost of cutting most of the galaxies in our sample with $V_{W20,i} < 100$ \kms (see Figure \ref{fig_btf_syst_all}, Panels (F) and (J)). Indeed, many resolved BTFR studies tend to select galaxies that have flat rotation curves and then discard galaxies that do not. 

A recent study by \citet{Brook:2015fz} illustrates the power and danger of various rotation velocity definitions for low-mass galaxies. In this work, the authors use the $M_{\rm baryon}$ to halo mass ($M_{\rm halo}$) abundance matching technique to calculate the total baryonic mass for simulated dark matter halos. Using several BTFRs in the literature, the authors convert $M_{\rm baryon}$ to various rotation velocity definitions. The authors show that depending on the velocity measurement used to calibrate the BTFR, the overabundance problem essentially disappears \citep[e.g.,][]{Klypin:2015kr}. If the rotation velocities of low-mass galaxies are poorly measured by \hi~line widths, \hi~line width functions may also be severely affected \citep{Maccio:2016tp}. Therefore, the definition of rotation velocity may create, or at the very least exacerbate, the overabundance problem found by, \citet[e.g.,][]{Papastergis:2011ie} and \citet{Klypin:2015kr}. 

Indeed, \citet{SantosSantos:2015em} have used 22 of such simulated galaxies to replicate the observed relation of \citet{McGaugh:2015eu}. These authors fit a slope of 3.5 to the maximum circular velocity BTFR while they fit a slope of 3.8 to the flattening velocity BTFR. \citet{Brook:2016gh} have also studied how various observed velocity measurements can affect the resulting BTFR using a sample of isolated simulated galaxies. These authors find that the low-mass end of the BTFR is especially susceptible to rotation velocity measurement. The authors examine different velocity measurements with mock \hi~line widths and \hi~rotation curves. They also impose the requirement that all galaxies have asymptotic rotation curves and find that their galaxies are consistent with the measurements of Paper~\citetalias{Bradford:2015km}, \citet{Lelli:2016ed} and \citet{DiCintio:2016tj}. They confirm that simulated low-mass galaxies fall off the $V_{\rm flat}$ BTFR relation due to rising rotation curves. 

In summary, we do not advocate for any one ``correct" rotation velocity definition for the BTFR, we only point out that results are heavily dependent on the details of the observations and we conclude that when comparing observed and predicted BTFRs, rotation velocity definitions must match in order for the comparison to be meaningful. Our results imply that existing line widths studies are all affected by a variety of systematics and these systematics are especially damaging for dynamical observations of low-mass galaxies. It appears that the ``true" BTFR using $V_{\rm flat}$ can only be measured using resolved, well-behaved \hi~rotation curves. Therefore any comparison to theoretical predictions of the BTFR must only include simulated or modeled galaxies where the \hi~gas has probed $V_{\rm flat}$ in the same, clearly defined way. This requires a full baryonic treatment and mock resolved \hi~observations that have had comparable quality cuts imposed on the mock observations. If our study of edge-on galaxies in systematic (F) and the recent study by \citet{Papastergis:2016tl} are any indication of the galaxy population that have observable asymptotic rotation curves within the unresolved samples, we are essentially forced to discard more than 90\% of our galaxy sample in order to obtain the correct dynamical measure of our galaxies using line widths. Nearly all of the galaxies with $V_{\rm rot} < 100$\kms~are discarded due to these cuts. In order to take advantage of the richness of these large single-dish data sets (HIPASS, ALFALFA and the upcoming SKA), mock unresolved observations should be implemented for models and simulations. Therefore the critical next steps are to calibrate the BTFR using these mock observations from a full hydrodynamical simulation and to follow-up with more resolved rotation curves for our low-mass sample.

Given these results and the fact that we have fit such a wide range of slopes to our data set, ruling out or confirming predictions of galaxy scaling relations or galaxy abundances in a $\Lambda$CDM framework takes more than simply comparing to observations if the parameters of the predictions are not well-matched to the observed galaxies. Finally, we note that information is lost when comparing the slope and scatter of various BTFRs. Instead, it is more productive to directly compare the distribution of galaxies with similar definitions of $M_{\rm baron}$ and $V_{\rm rot}$, as opposed to comparing the fits to these distributions \citep[e.g.,][]{Hogg:2010wk}. The richness of data can be lost when we fit simple models to the BTFR. 

\begin{acknowledgements}
\textit{Dedicated to the memory of Janet C. Bradford}. We thank the anonymous referee and Stacy McGaugh for helpful comments that have improved the quality of this manuscript. We would like to thank Stephane Courteau, Susan Kassin, Allison Merritt, and Erik Tollerud for helpful feedback on this manuscript, as well as Michael Blanton and the BS group for helpful discussions. J.D.B. acknowledges support from the Gruber Foundation and the National Science Foundation Graduate Research Fellowship Program. M.G. acknowledges a fellowship from the John S. Guggenheim Memorial Foundation. The Arecibo Observatory is operated by SRI International under a cooperative agreement with the National Science Foundation (AST-1100968), and in alliance with Ana G. M\'{e}ndez-Universidad Metropolitana, and the Universities Space Research Association. The National Radio Astronomy Observatory is a facility of the National Science Foundation operated under cooperative agreement by Associated Universities, Inc. Funding for SDSS-III has been provided by the Alfred P. Sloan Foundation, the Participating Institutions, the National Science Foundation, and the U.S. Department of Energy Office of Science. The SDSS-III web site is http://www.sdss3.org/. This material is based upon work supported by the National Science Foundation Graduate Research Fellowship Program under Grant No. DGE-1122492. Any opinions, findings, and conclusions or recommendations expressed in this material are those of the author(s) and do not necessarily reflect the views of the National Science Foundation.

\end{acknowledgements}


\end{document}